# Boosting Algorithms: Regularization, Prediction and Model Fitting

**Peter Bühlmann and Torsten Hothorn**

*Abstract.* We present a statistical perspective on boosting. Special emphasis is given to estimating potentially complex parametric or nonparametric models, including generalized linear and additive models as well as regression models for survival analysis. Concepts of degrees of freedom and corresponding Akaike or Bayesian information criteria, particularly useful for regularization and variable selection in high-dimensional covariate spaces, are discussed as well.

The practical aspects of boosting procedures for fitting statistical models are illustrated by means of the dedicated open-source software package **mboost**. This package implements functions which can be used for model fitting, prediction and variable selection. It is flexible, allowing for the implementation of new boosting algorithms optimizing user-specified loss functions.

## 1. INTRODUCTION

Freund and Schapire's AdaBoost algorithm for classification **(author?)** [29, 30, 31] has attracted much attention in the machine learning community (cf. [76], and the references therein) as well as in related areas in statistics **(author?)** [15, 16, 33]. Various versions of the AdaBoost algorithm have proven to be very competitive in terms of prediction accuracy in a variety of applications. Boosting methods have been originally proposed as ensemble methods (see Section 1.1), which rely on the principle of generating multiple predictions and majority voting (averaging) among the individual classifiers.

Later, Breiman **(author?)** [15, 16] made a path-breaking observation that the AdaBoost algorithm can be viewed as a gradient descent algorithm in function space, inspired by numerical optimization and statistical estimation. Moreover, Friedman, Hastie and Tibshirani **(author?)** [33] laid out further important foundations which linked Ada-Boost and other boosting algorithms to the framework of statistical estimation and additive basis expansion. In their terminology, boosting is represented as "stagewise, additive modeling": the word "additive" does not imply a model fit which is additive in the covariates (see our Section 4), but refers to the fact that boosting is an additive (in fact, a linear) combination of "simple" (function) estimators. Also Mason et al. **(author?)** [62] and Rätsch, Onoda and Müller **(author?)** [70] developed related ideas which were mainly acknowledged in the machine learning community. In Hastie, Tibshirani and Friedman **(author?)** [42], additional views on boosting are given;

*Peter Bühlmann is Professor, Seminar für Statistik, ETH Zürich, CH-8092 Zürich, Switzerland e-mail: buhlmann@stat.math.ethz.ch. Torsten Hothorn is Professor, Institut für Statistik, Ludwig-Maximilians-Universität München, Ludwigstraße 33, D-80539 München, Germany e-mail: Torsten.Hothorn@R-project.org. Torsten Hothorn wrote this paper while he was a lecturer at the Universität Erlangen-Nürnberg.*

[1]Discussed in 10.1214/07-STS242A and 10.1214/07-STS242B; rejoinder at 10.1214/07-STS242REJ.

in particular, the authors first pointed out the relation between boosting and $\ell^1$-penalized estimation. The insights of Friedman, Hastie and Tibshirani **(author?)** [33] opened new perspectives, namely to use boosting methods in many other contexts than classification. We mention here boosting methods for regression (including generalized regression) [22, 32, 71], for density estimation [73], for survival analysis [45, 71] or for multivariate analysis [33, 59]. In quite a few of these proposals, boosting is not only a black-box prediction tool but also an estimation method for models with a specific structure such as linearity or additivity [18, 22, 45]. Boosting can then be seen as an interesting regularization scheme for estimating a model. This statistical perspective will drive the focus of our exposition of boosting.

We present here some coherent explanations and illustrations of concepts about boosting, some derivations which are novel, and we aim to increase the understanding of some methods and some selected known results. Besides giving an overview on theoretical concepts of boosting as an algorithm for fitting statistical models, we look at the methodology from a practical point of view as well. The dedicated add-on package **mboost** ("model-based boosting," [43]) to the R system for statistical computing [69] implements computational tools which enable the data analyst to compute on the theoretical concepts explained in this paper as closely as possible. The illustrations presented throughout the paper focus on three regression problems with continuous, binary and censored response variables, some of them having a large number of covariates. For each example, we only present the most important steps of the analysis. The complete analysis is contained in a vignette as part of the **mboost** package (see Appendix A.1) so that every result shown in this paper is reproducible.

Unless stated differently, we assume that the data are realizations of random variables

$$(X_1, Y_1), \ldots, (X_n, Y_n)$$

from a stationary process with $p$-dimensional predictor variables $X_i$ and one-dimensional response variables $Y_i$; for the case of multivariate responses, some references are given in Section 9.1. In particular, the setting above includes independent, identically distributed (i.i.d.) observations. The generalization to stationary processes is fairly straightforward: the methods and algorithms are the same as in the i.i.d. framework, but the mathematical theory requires more elaborate techniques. Essentially, one needs to ensure that some (uniform) laws of large numbers still hold, for example, assuming stationary, mixing sequences; some rigorous results are given in [57] and [59].

### 1.1 Ensemble Schemes: Multiple Prediction and Aggregation

Ensemble schemes construct multiple function estimates or predictions from reweighted data and use a linear (or sometimes convex) combination thereof for producing the final, aggregated estimator or prediction.

First, we specify a *base procedure* which constructs a function estimate $\hat{g}(\cdot)$ with values in $\mathbb{R}$, based on some data $(X_1, Y_1), \ldots, (X_n, Y_n)$:

$$(X_1, Y_1), \ldots, (X_n, Y_n) \xrightarrow{\text{base procedure}} \hat{g}(\cdot).$$

For example, a very popular base procedure is a regression tree.

Then, generating an ensemble from the base procedures, that is, an ensemble of function estimates or predictions, works generally as follows:

$$\begin{aligned}
\text{reweighted data 1} &\xrightarrow{\text{base procedure}} \hat{g}^{[1]}(\cdot) \\
\text{reweighted data 2} &\xrightarrow{\text{base procedure}} \hat{g}^{[2]}(\cdot) \\
\cdots & \quad \cdots \\
\cdots & \quad \cdots \\
\text{reweighted data } M &\xrightarrow{\text{base procedure}} \hat{g}^{[M]}(\cdot) \\
\text{aggregation: } \hat{f}_A(\cdot) &= \sum_{m=1}^{M} \alpha_m \hat{g}^{[m]}(\cdot).
\end{aligned}$$

What is termed here as "reweighted data" means that we assign individual data weights to each of the $n$ sample points. We have also implicitly assumed that the base procedure allows to do some weighted fitting, that is, estimation is based on a weighted sample. Throughout the paper (except in Section 1.2), we assume that a base procedure estimate $\hat{g}(\cdot)$ is real-valued (i.e., a regression procedure), making it more adequate for the "statistical perspective" on boosting, in particular for the generic FGD algorithm in Section 2.1.

The above description of an ensemble scheme is too general to be of any direct use. The specification of the data reweighting mechanism as well as the form of the linear combination coefficients $\{\alpha_m\}_{m=1}^{M}$ are crucial, and various choices characterize different ensemble schemes. Most boosting methods are special kinds of *sequential* ensemble schemes, where the data weights in iteration $m$ depend on the results from the previous iteration $m-1$ only (*memoryless* with respect to iterations $m-2, m-3, \ldots$). Examples of other ensemble schemes include bagging [14] or random forests [1, 17].

## 1.2 AdaBoost

The AdaBoost algorithm for binary classification [31] is the most well-known boosting algorithm. The base procedure is a classifier with values in $\{0,1\}$ (slightly different from a real-valued function estimator as assumed above), for example, a classification tree.

### AdaBoost algorithm

1. Initialize some weights for individual sample points: $w_i^{[0]} = 1/n$ for $i = 1, \ldots, n$. Set $m = 0$.
2. Increase $m$ by 1. Fit the base procedure to the weighted data, that is, do a weighted fitting using the weights $w_i^{[m-1]}$, yielding the classifier $\hat{g}^{[m]}(\cdot)$.
3. Compute the weighted in-sample misclassification rate

$$\mathrm{err}^{[m]} = \sum_{i=1}^{n} w_i^{[m-1]} I(Y_i \neq \hat{g}^{[m]}(X_i)) \Big/ \sum_{i=1}^{n} w_i^{[m-1]},$$

$$\alpha^{[m]} = \log\left(\frac{1 - \mathrm{err}^{[m]}}{\mathrm{err}^{[m]}}\right),$$

and update the weights

$$\tilde{w}_i = w_i^{[m-1]} \exp(\alpha^{[m]} I(Y_i \neq \hat{g}^{[m]}(X_i))),$$

$$w_i^{[m]} = \tilde{w}_i \Big/ \sum_{j=1}^{n} \tilde{w}_j.$$

4. Iterate steps 2 and 3 until $m = m_{\mathrm{stop}}$ and build the aggregated classifier by weighted majority voting:

$$\hat{f}_{\mathrm{AdaBoost}}(x) = \arg\max_{y \in \{0,1\}} \sum_{m=1}^{m_{\mathrm{stop}}} \alpha^{[m]} I(\hat{g}^{[m]}(x) = y).$$

By using the terminology $m_{\mathrm{stop}}$ (instead of $M$ as in the general description of ensemble schemes), we emphasize here and later that the iteration process should be stopped to avoid overfitting. It is a tuning parameter of AdaBoost which may be selected using some cross-validation scheme.

## 1.3 Slow Overfitting Behavior

It had been debated until about the year 2000 whether the AdaBoost algorithm is immune to overfitting when running more iterations, that is, stopping would not be necessary. It is clear nowadays that Ada-Boost and also other boosting algorithms are overfitting eventually, and early stopping [using a value of $m_{\mathrm{stop}}$ before convergence of the surrogate loss function, given in (3.3), takes place] is necessary [7, 51, 64]. We emphasize that this is not in contradiction to the experimental results by **(author?)** [15] where the test set misclassification error still decreases after the training misclassification error is zero [because the training error of the surrogate loss function in (3.3) is not zero before numerical convergence].

Nevertheless, the AdaBoost algorithm is quite resistant to overfitting (slow overfitting behavior) when increasing the number of iterations $m_{\mathrm{stop}}$. This has been observed empirically, although some cases with clear overfitting do occur for some datasets [64]. A stream of work has been devoted to develop VC-type bounds for the generalization (out-of-sample) error to explain why boosting is overfitting very slowly only. Schapire et al. **(author?)** [77] proved a remarkable bound for the generalization misclassification error for classifiers in the convex hull of a base procedure. This bound for the misclassification error has been improved by Koltchinskii and Panchenko **(author?)** [53], deriving also a generalization bound for AdaBoost which depends on the number of boosting iterations.

It has been argued in [33], rejoinder, and [21] that the overfitting resistance (slow overfitting behavior) is much stronger for the misclassification error than many other loss functions such as the (out-of-sample) negative log-likelihood (e.g., squared error in Gaussian regression). Thus, boosting's resistance of overfitting is coupled with a general fact that overfitting is less an issue for classification (i.e., the 0-1 loss function). Furthermore, it is proved in [6] that the misclassification risk can be bounded by the risk of the surrogate loss function: it demonstrates from a different perspective that the 0-1 loss can exhibit quite a different behavior than the surrogate loss.

Finally, Section 5.1 develops the variance and bias for boosting when utilized to fit a one-dimensional curve. Figure 5 illustrates the difference between the boosting and the smoothing spline approach, and the eigen-analysis of the boosting method [see (5.2)] yields the following: boosting's variance increases with exponentially small increments while its squared bias decreases exponentially fast as the number of iterations grows. This also explains why boosting's overfitting kicks in very slowly.

### 1.4 Historical Remarks

The idea of boosting as an ensemble method for improving the predictive performance of a base procedure seems to have its roots in machine learning. Kearns and Valiant **(author?)** [52] proved that if individual classifiers perform at least slightly better than guessing at random, their predictions can be combined and averaged, yielding much better predictions. Later, Schapire **(author?)** [75] proposed a boosting algorithm with provable polynomial runtime to construct such a better ensemble of classifiers. The AdaBoost algorithm [29, 30, 31] is considered as a first path-breaking step toward practically feasible boosting algorithms.

The results from Breiman **(author?)** [15, 16], showing that boosting can be interpreted as a functional gradient descent algorithm, uncover older roots of boosting. In the context of regression, there is an immediate connection to the Gauss–Southwell algorithm [79] for solving a linear system of equations (see Section 4.1) and to Tukey's [83] method of "twicing" (see Section 5.1).

## 2. FUNCTIONAL GRADIENT DESCENT

Breiman **(author?)** [15, 16] showed that the AdaBoost algorithm can be represented as a steepest descent algorithm in function space which we call functional gradient descent (FGD). Friedman, Hastie and Tibshirani **(author?)** [33] and Friedman **(author?)** [32] then developed a more general, statistical framework which yields a direct interpretation of boosting as a method for function estimation. In their terminology, it is a "stagewise, additive modeling" approach (but the word "additive" does not imply a model fit which is additive in the covariates; see Section 4). Consider the problem of estimating a real-valued function

$$(2.1) \qquad f^*(\cdot) = \arg\min_{f(\cdot)} \mathbb{E}[\rho(Y, f(X))],$$

where $\rho(\cdot, \cdot)$ is a loss function which is typically assumed to be differentiable and convex with respect to the second argument. For example, the squared error loss $\rho(y, f) = |y - f|^2$ yields the well-known population minimizer $f^*(x) = \mathbb{E}[Y|X = x]$.

### 2.1 The Generic FGD or Boosting Algorithm

In the sequel, FGD and boosting are used as equivalent terminology for the same method or algorithm.

Estimation of $f^*(\cdot)$ in (2.1) with boosting can be done by considering the empirical risk $n^{-1} \sum_{i=1}^n \rho(Y_i, f(X_i))$ and pursuing iterative steepest descent in function space. The following algorithm has been given by Friedman **(author?)** [32]:

Generic FGD algorithm

1. Initialize $\hat{f}^{[0]}(\cdot)$ with an offset value. Common choices are

$$\hat{f}^{[0]}(\cdot) \equiv \arg\min_c n^{-1} \sum_{i=1}^n \rho(Y_i, c)$$

or $\hat{f}^{[0]}(\cdot) \equiv 0$. Set $m = 0$.

2. Increase $m$ by 1. Compute the negative gradient $-\frac{\partial}{\partial f}\rho(Y, f)$ and evaluate at $\hat{f}^{[m-1]}(X_i)$:

$$U_i = -\frac{\partial}{\partial f}\rho(Y_i, f)|_{f=\hat{f}^{[m-1]}(X_i)}, \quad i = 1, \ldots, n.$$

3. Fit the negative gradient vector $U_1, \ldots, U_n$ to $X_1, \ldots, X_n$ by the real-valued base procedure (e.g., regression)

$$(X_i, U_i)_{i=1}^n \quad \xrightarrow{\text{base procedure}} \quad \hat{g}^{[m]}(\cdot).$$

Thus, $\hat{g}^{[m]}(\cdot)$ can be viewed as an approximation of the negative gradient vector.

4. Update $\hat{f}^{[m]}(\cdot) = \hat{f}^{[m-1]}(\cdot) + \nu \cdot \hat{g}^{[m]}(\cdot)$, where $0 < \nu \leq 1$ is a step-length factor (see below), that is, proceed along an estimate of the negative gradient vector.

5. Iterate steps 2 to 4 until $m = m_{\text{stop}}$ for some stopping iteration $m_{\text{stop}}$.

The stopping iteration, which is the main tuning parameter, can be determined via cross-validation or some information criterion; see Section 5.4. The choice of the step-length factor $\nu$ in step 4 is of minor importance, as long as it is "small," such as $\nu = 0.1$. A smaller value of $\nu$ typically requires a larger number of boosting iterations and thus more computing time, while the predictive accuracy has been empirically found to be potentially better and almost never worse when choosing $\nu$ "sufficiently small" (e.g., $\nu = 0.1$) [32]. Friedman **(author?)** [32] suggests to use an additional line search between steps 3 and 4 (in case of other loss functions $\rho(\cdot, \cdot)$ than squared error): it yields a slightly different algorithm but the additional line search seems unnecessary for achieving a good estimator $\hat{f}^{[m_{\text{stop}}]}$. The latter statement is based on empirical evidence and some mathematical reasoning as described at the beginning of Section 7.

2.1.1 *Alternative formulation in function space.*
In steps 2 and 3 of the generic FGD algorithm, we associated with $U_1,\ldots,U_n$ a negative gradient vector. A reason for this can be seen from the following formulation in function space which is similar to the exposition in Mason et al. **(author?)** [62] and to the discussion in Ridgeway **(author?)** [72].

Consider the empirical risk functional $C(f) = n^{-1}\sum_{i=1}^n \rho(Y_i, f(X_i))$ and the usual inner product $\langle f, g \rangle = n^{-1}\sum_{i=1}^n f(X_i)g(X_i)$. We can then calculate the negative Gâteaux derivative $dC(\cdot)$ of the functional $C(\cdot)$,

$$-dC(f)(x) = -\frac{\partial}{\partial \alpha} C(f + \alpha \delta_x)|_{\alpha=0},$$

$$f : \mathbb{R}^p \to \mathbb{R},\ x \in \mathbb{R}^p,$$

where $\delta_x$ denotes the delta- (or indicator-) function at $x \in \mathbb{R}^p$. In particular, when evaluating the derivative $-dC$ at $\hat{f}^{[m-1]}$ and $X_i$, we get

$$-dC(\hat{f}^{[m-1]})(X_i) = n^{-1} U_i,$$

with $U_1,\ldots,U_n$ exactly as in steps 2 and 3 of the generic FGD algorithm. Thus, the negative gradient vector $U_1,\ldots,U_n$ can be interpreted as a functional (Gâteaux) derivative evaluated at the data points.

We point out that the algorithm in Mason et al. **(author?)** [62] is different from the generic FGD method above: while the latter is fitting the negative gradient vector by the base procedure, typically using (nonparametric) least squares, Mason et al. **(author?)** [62] fit the base procedure by maximizing $-\langle U, \hat{g} \rangle = n^{-1}\sum_{i=1}^n U_i \hat{g}(X_i)$. For certain base procedures, the two algorithms coincide. For example, if $\hat{g}(\cdot)$ is the componentwise linear least squares base procedure described in (4.1), it holds that $n^{-1}\sum_{i=1}^n (U_i - \hat{g}(X_i))^2 = C - \langle U, \hat{g} \rangle$, where $C = n^{-1}\sum_{i=1}^n U_i^2$ is a constant.

## 3. SOME LOSS FUNCTIONS AND BOOSTING ALGORITHMS

Various boosting algorithms can be defined by specifying different (surrogate) loss functions $\rho(\cdot, \cdot)$. The **mboost** package provides an environment for defining loss functions via *boost_family* objects, as exemplified below.

### 3.1 Binary Classification

For binary classification, the response variable is $Y \in \{0, 1\}$ with $\mathbb{P}[Y = 1] = p$. Often, it is notationally more convenient to encode the response by $\tilde{Y} = 2Y - 1 \in \{-1, +1\}$ (this coding is used in **mboost** as well). We consider the negative binomial log-likelihood as loss function:

$$-(y\log(p) + (1-y)\log(1-p)).$$

We parametrize $p = \exp(f)/(\exp(f) + \exp(-f))$ so that $f = \log(p/(1-p))/2$ equals half of the log-odds ratio; the factor $1/2$ is a bit unusual but it will enable that the population minimizer of the loss in (3.1) is the same as for the exponential loss in (3.3) below. Then, the negative log-likelihood is

$$\log(1 + \exp(-2\tilde{y}f)).$$

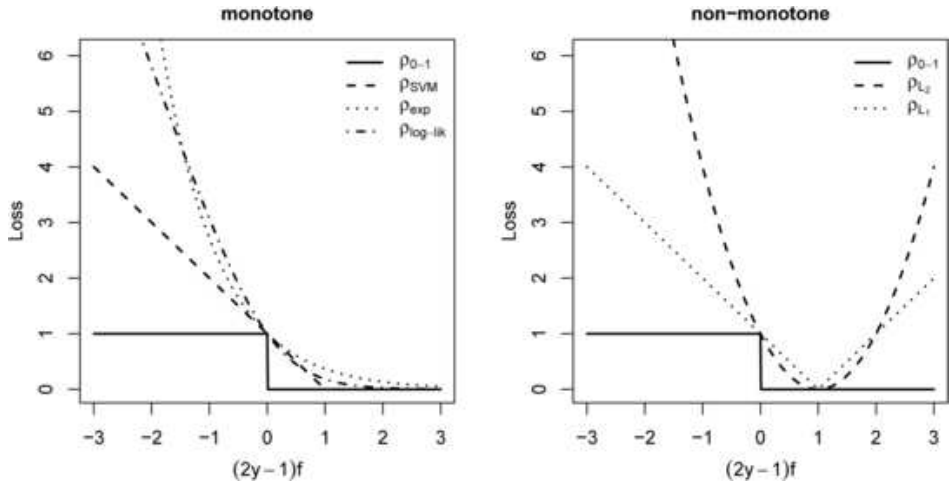

Fig. 1. *Losses, as functions of the margin $\tilde{y}f = (2y-1)f$, for binary classification. Left panel with monotone loss functions: 0-1 loss, exponential loss, negative log-likelihood, hinge loss (SVM); right panel with nonmonotone loss functions: squared error ($L_2$) and absolute error ($L_1$) as in (3.5).*

By scaling, we prefer to use the equivalent loss function

$$(3.1) \quad \rho_{\text{log-lik}}(\tilde{y}, f) = \log_2(1 + \exp(-2\tilde{y}f)),$$

which then becomes an upper bound of the misclassification error; see Figure 1. In **mboost**, the negative gradient of this loss function is implemented in a function `Binomial()` returning an object of class *boost_family* which contains the negative gradient *function* as a slot (assuming a binary response variable y ∈ {−1, +1}).

The population minimizer can be shown to be (cf. [33])

$$f^*_{\text{log-lik}}(x) = \frac{1}{2} \log\left(\frac{p(x)}{1 - p(x)}\right),$$
$$p(x) = \mathbb{P}[Y = 1 | X = x].$$

The loss function in (3.1) is a function of $\tilde{y}f$, the so-called margin value, where the function $f$ induces the following classifier for $Y$:

$$\mathcal{C}(x) = \begin{cases} 1, & \text{if } f(x) > 0, \\ 0, & \text{if } f(x) < 0, \\ \text{undetermined}, & \text{if } f(x) = 0. \end{cases}$$

Therefore, a misclassification (including the undetermined case) happens if and only if $\tilde{Y}f(X) \leq 0$. Hence, the misclassification loss is

$$(3.2) \qquad \rho_{0-1}(y, f) = I_{\{\tilde{y}f \leq 0\}},$$

whose population minimizer is equivalent to the Bayes classifier (for $\tilde{Y} \in \{-1, +1\}$)

$$f^*_{0-1}(x) = \begin{cases} +1, & \text{if } p(x) > 1/2, \\ -1, & \text{if } p(x) \leq 1/2, \end{cases}$$

where $p(x) = \mathbb{P}[Y = 1 | X = x]$. Note that the 0-1 loss in (3.2) cannot be used for boosting or FGD: it is nondifferentiable and also nonconvex as a function of the margin value $\tilde{y}f$. The negative log-likelihood loss in (3.1) can be viewed as a convex upper approximation of the (computationally intractable) nonconvex 0-1 loss; see Figure 1. We will describe in Section 3.3 the BinomialBoosting algorithm (similar to LogitBoost [33]) which uses the negative log-likelihood as loss function (i.e., the surrogate loss which is the implementing loss function for the algorithm).

Another upper convex approximation of the 0-1 loss function in (3.2) is the exponential loss

$$(3.3) \qquad \rho_{\exp}(y, f) = \exp(-\tilde{y}f),$$

implemented (with notation y ∈ {−1, +1}) in **mboost** as `AdaExp()` family.

The population minimizer can be shown to be the same as for the log-likelihood loss (cf. [33]):

$$f^*_{\exp}(x) = \frac{1}{2} \log\left(\frac{p(x)}{1 - p(x)}\right),$$
$$p(x) = \mathbb{P}[Y = 1 | X = x].$$

Using functional gradient descent with different (surrogate) loss functions yields different boosting algorithms. When using the log-likelihood loss in (3.1), we obtain LogitBoost [33] or BinomialBoosting from Section 3.3; and with the exponential loss in (3.3), we essentially get AdaBoost [30] from Section 1.2.

We interpret the boosting estimate $\hat{f}^{[m]}(\cdot)$ as an estimate of the population minimizer $f^*(\cdot)$. Thus, the output from AdaBoost, Logit- or BinomialBoosting are estimates of half of the log-odds ratio. In particular, we define probability estimates via

$$\hat{p}^{[m]}(x) = \frac{\exp(\hat{f}^{[m]}(x))}{\exp(\hat{f}^{[m]}(x)) + \exp(-\hat{f}^{[m]}(x))}.$$

The reason for constructing these probability estimates is based on the fact that boosting with a suitable stopping iteration is consistent [7, 51]. Some cautionary remarks about this line of argumentation are presented by Mease, Wyner and Buja **(author?)** [64].

Very popular in machine learning is the hinge function, the standard loss function for support vector machines:

$$\rho_{\text{SVM}}(y, f) = [1 - \tilde{y}f]_+,$$

where $[x]_+ = xI_{\{x>0\}}$ denotes the positive part. It is also an upper convex bound of the misclassification error; see Figure 1. Its population minimizer is

$$f^*_{\text{SVM}}(x) = \text{sign}(p(x) - 1/2),$$

which is the Bayes classifier for $\tilde{Y} \in \{-1, +1\}$. Since $f^*_{\text{SVM}}(\cdot)$ is a classifier and noninvertible function of $p(x)$, there is no direct way to obtain conditional class probability estimates.

### 3.2 Regression

For regression with response $Y \in \mathbb{R}$, we use most often the squared error loss (scaled by the factor 1/2 such that the negative gradient vector equals the residuals; see Section 3.3 below),

$$(3.4) \qquad \rho_{L_2}(y, f) = \tfrac{1}{2}|y - f|^2$$

with population minimizer

$$f_{L_2}^*(x) = \mathbb{E}[Y|X = x].$$

The corresponding boosting algorithm is $L_2$Boosting; see Friedman **(author?)** [32] and Bühlmann and Yu **(author?)** [22]. It is described in more detail in Section 3.3. This loss function is available in **mboost** as family `GaussReg()`.

Alternative loss functions which have some robustness properties (with respect to the error distribution, i.e., in "Y-space") include the $L_1$- and Huber-loss. The former is

$$\rho_{L_1}(y, f) = |y - f|$$

with population minimizer

$$f^*(x) = \text{median}(Y|X = x)$$

and is implemented in **mboost** as `Laplace()`.

Although the $L_1$-loss is not differentiable at the point $y = f$, we can compute partial derivatives since the single point $y = f$ (usually) has probability zero to be realized by the data. A compromise between the $L_1$- and $L_2$-loss is the Huber-loss function from robust statistics:

$$\rho_{\text{Huber}}(y, f)$$
$$= \begin{cases} |y - f|^2/2, & \text{if } |y - f| \leq \delta, \\ \delta(|y - f| - \delta/2), & \text{if } |y - f| > \delta, \end{cases}$$

which is available in **mboost** as `Huber()`. A strategy for choosing (a changing) $\delta$ adaptively has been proposed by Friedman **(author?)** [32]:

$$\delta_m = \text{median}(\{|Y_i - \hat{f}^{[m-1]}(X_i)|; i = 1, \ldots, n\}),$$

where the previous fit $\hat{f}^{[m-1]}(\cdot)$ is used.

3.2.1 *Connections to binary classification.* Motivated from the population point of view, the $L_2$- or $L_1$-loss can also be used for binary classification. For $Y \in \{0, 1\}$, the population minimizers are

$$f_{L_2}^*(x) = \mathbb{E}[Y|X = x]$$
$$= p(x) = \mathbb{P}[Y = 1|X = x],$$
$$f_{L_1}^*(x) = \text{median}(Y|X = x)$$
$$= \begin{cases} 1, & \text{if } p(x) > 1/2, \\ 0, & \text{if } p(x) \leq 1/2. \end{cases}$$

Thus, the population minimizer of the $L_1$-loss is the Bayes classifier.

Moreover, both the $L_1$- and $L_2$-loss functions can be parametrized as functions of the margin value $\tilde{y}f$ ($\tilde{y} \in \{-1, +1\}$):

$$|\tilde{y} - f| = |1 - \tilde{y}f|,$$
(3.5) $$|\tilde{y} - f|^2 = |1 - \tilde{y}f|^2$$
$$= (1 - 2\tilde{y}f + (\tilde{y}f)^2).$$

The $L_1$- and $L_2$-loss functions are nonmonotone functions of the margin value $\tilde{y}f$; see Figure 1. A negative aspect is that they penalize margin values which are greater than 1: penalizing large margin values can be seen as a way to encourage solutions $\hat{f} \in [-1, 1]$ which is the range of the population minimizers $f_{L_1}^*$ and $f_{L_2}^*$ (for $\tilde{Y} \in \{-1, +1\}$), respectively. However, as discussed below, we prefer to use monotone loss functions.

The $L_2$-loss for classification (with response variable $\texttt{y} \in \{-1, +1\}$) is implemented in `GaussClass()`.

All loss functions mentioned for binary classification (displayed in Figure 1) can be viewed and interpreted from the perspective of proper scoring rules; cf. Buja, Stuetzle and Shen **(author?)** [24]. We usually prefer the negative log-likelihood loss in (3.1) because: (i) it yields probability estimates; (ii) it is a monotone loss function of the margin value $\tilde{y}f$; (iii) it grows linearly as the margin value $\tilde{y}f$ tends to $-\infty$, unlike the exponential loss in (3.3). The third point reflects a robustness aspect: it is similar to Huber's loss function which also penalizes large values linearly (instead of quadratically as with the $L_2$-loss).

### 3.3 Two Important Boosting Algorithms

Table 1 summarizes the most popular loss functions and their corresponding boosting algorithms. We now describe the two algorithms appearing in the last two rows of Table 1 in more detail.

3.3.1 $L_2$*Boosting.* $L_2$Boosting is the simplest and perhaps most instructive boosting algorithm. It is very useful for regression, in particular in presence of very many predictor variables. Applying the general description of the FGD algorithm from Section 2.1 to the squared error loss function $\rho_{L_2}(y, f) = |y - f|^2/2$, we obtain the following algorithm:

$L_2$Boosting algorithm

1. Initialize $\hat{f}^{[0]}(\cdot)$ with an offset value. The default value is $\hat{f}^{[0]}(\cdot) \equiv \overline{Y}$. Set $m = 0$.
2. Increase $m$ by 1. Compute the residuals $U_i = Y_i - \hat{f}^{[m-1]}(X_i)$ for $i = 1, \ldots, n$.

3. Fit the residual vector $U_1, \ldots, U_n$ to $X_1, \ldots, X_n$ by the real-valued base procedure (e.g., regression):

$$(X_i, U_i)_{i=1}^n \xrightarrow{\text{base procedure}} \hat{g}^{[m]}(\cdot).$$

4. Update $\hat{f}^{[m]}(\cdot) = \hat{f}^{[m-1]}(\cdot) + \nu \cdot \hat{g}^{[m]}(\cdot)$, where $0 < \nu \leq 1$ is a step-length factor (as in the general FGD algorithm).
5. Iterate steps 2 to 4 until $m = m_{\text{stop}}$ for some stopping iteration $m_{\text{stop}}$.

The stopping iteration $m_{\text{stop}}$ is the main tuning parameter which can be selected using cross-validation or some information criterion as described in Section 5.4.

The derivation from the generic FGD algorithm in Section 2.1 is straightforward. Note that the negative gradient vector becomes the residual vector. Thus, $L_2$Boosting amounts to refitting residuals multiple times. Tukey **(author?)** [83] recognized this to be useful and proposed "twicing," which is nothing else than $L_2$Boosting using $m_{\text{stop}} = 2$ (and $\nu = 1$).

3.3.2 *BinomialBoosting: the FGD version of Logit-Boost.* We already gave some reasons at the end of Section 3.2.1 why the negative log-likelihood loss function in (3.1) is very useful for binary classification problems. Friedman, Hastie and Tibshirani **(author?)** [33] were first in advocating this, and they proposed Logit-Boost, which is very similar to the generic FGD algorithm when using the loss from (3.1): the deviation from FGD is the use of Newton's method involving the Hessian matrix (instead of a step-length for the gradient).

For the sake of coherence with the generic functional gradient descent algorithm in Section 2.1, we describe here a version of LogitBoost; to avoid conflicting terminology, we name it BinomialBoosting:

BinomialBoosting algorithm

Apply the generic FGD algorithm from Section 2.1 using the loss function $\rho_{\text{log-lik}}$ from (3.1). The default offset value is $\hat{f}^{[0]}(\cdot) \equiv \log(\hat{p}/(1 - \hat{p}))/2$, where $\hat{p}$ is the relative frequency of $Y = 1$.

With BinomialBoosting, there is no need that the base procedure is able to do weighted fitting; this constitutes a slight difference to the requirement for Logit-Boost [33].

### 3.4 Other Data Structures and Models

Due to the generic nature of boosting or functional gradient descent, we can use the technique in very many other settings. For data with univariate responses and loss functions which are differentiable with respect to the second argument, the boosting algorithm is described in Section 2.1. Survival analysis is an important area of application with censored observations; we describe in Section 8 how to deal with it.

## 4. CHOOSING THE BASE PROCEDURE

Every boosting algorithm requires the specification of a base procedure. This choice can be driven by the aim of optimizing the predictive capacity only or by considering some structural properties of the boosting estimate in addition. We find the latter usually more interesting as it allows for better interpretation of the resulting model.

We recall that the generic boosting estimator is a sum of base procedure estimates

$$\hat{f}^{[m]}(\cdot) = \nu \sum_{k=1}^{m} \hat{g}^{[k]}(\cdot).$$

Therefore, structural properties of the boosting function estimator are induced by a linear combination of structural characteristics of the base procedure.

The following important examples of base procedures yield useful structures for the boosting estimator $\hat{f}^{[m]}(\cdot)$. The notation is as follows: $\hat{g}(\cdot)$ is an estimate from a base procedure which is based on data $(X_1, U_1), \ldots, (X_n, U_n)$ where $(U_1, \ldots, U_n)$ denotes the current negative gradient. In the sequel, the $j$th component of a vector $c$ will be denoted by $c^{(j)}$.

TABLE 1
*Various loss functions $\rho(y, f)$, population minimizers $f^*(x)$ and names of corresponding boosting algorithms; $p(x) = \mathbb{P}[Y = 1 | X = x]$*

| Range spaces | $\rho(y, f)$ | $f^*(x)$ | Algorithm |
| --- | --- | --- | --- |
| $y \in \{0, 1\}, \ f \in \mathbb{R}$ | $\exp(-(2y-1)f)$ | $\frac{1}{2}\log(\frac{p(x)}{1-p(x)})$ | AdaBoost |
| $y \in \{0, 1\}, \ f \in \mathbb{R}$ | $\log_2(1 + e^{-2(2y-1)f})$ | $\frac{1}{2}\log(\frac{p(x)}{1-p(x)})$ | LogitBoost / BinomialBoosting |
| $y \in \mathbb{R}, \ f \in \mathbb{R}$ | $\frac{1}{2}|y - f|^2$ | $\mathbb{E}[Y|X=x]$ | $L_2$Boosting |

### 4.1 Componentwise Linear Least Squares for Linear Models

Boosting can be very useful for fitting potentially high-dimensional generalized linear models. Consider the base procedure

$$\hat{g}(x) = \hat{\beta}^{(\hat{\mathcal{S}})} x^{(\hat{\mathcal{S}})},$$

(4.1) $$\hat{\beta}^{(j)} = \sum_{i=1}^{n} X_i^{(j)} U_i \Big/ \sum_{i=1}^{n} (X_i^{(j)})^2,$$

$$\hat{\mathcal{S}} = \arg\min_{1 \leq j \leq p} \sum_{i=1}^{n} (U_i - \hat{\beta}^{(j)} X_i^{(j)})^2.$$

It selects the best variable in a simple linear model in the sense of ordinary least squares fitting.

When using $L_2$Boosting with this base procedure, we select in every iteration one predictor variable, not necessarily a different one for each iteration, and we update the function linearly:

$$\hat{f}^{[m]}(x) = \hat{f}^{[m-1]}(x) + \nu \hat{\beta}^{(\hat{\mathcal{S}}_m)} x^{(\hat{\mathcal{S}}_m)},$$

where $\hat{\mathcal{S}}_m$ denotes the index of the selected predictor variable in iteration $m$. Alternatively, the update of the coefficient estimates is

$$\hat{\beta}^{[m]} = \hat{\beta}^{[m-1]} + \nu \cdot \hat{\beta}^{(\hat{\mathcal{S}}_m)}.$$

The notation should be read that only the $\hat{\mathcal{S}}_m$th component of the coefficient estimate $\hat{\beta}^{[m]}$ (in iteration $m$) has been updated. For every iteration $m$, we obtain a linear model fit. As $m$ tends to infinity, $\hat{f}^{[m]}(\cdot)$ converges to a least squares solution which is unique if the design matrix has full rank $p \leq n$. The method is also known as matching pursuit in signal processing [60], weak greedy algorithm in computational mathematics [81], and it is a Gauss–Southwell algorithm [79] for solving a linear system of equations. We will discuss more properties of $L_2$Boosting with componentwise linear least squares in Section 5.2.

When using BinomialBoosting with componentwise linear least squares from (4.1), we obtain a fit, including variable selection, of a linear logistic regression model.

As will be discussed in more detail in Section 5.2, boosting typically shrinks the (logistic) regression coefficients toward zero. Usually, we do not want to shrink the intercept term. In addition, we advocate to use boosting on mean centered predictor variables $\tilde{X}_i^{(j)} = X_i^{(j)} - \overline{X}^{(j)}$. In case of a linear model, when centering also the response $\tilde{Y}_i = Y_i - \overline{Y}$, this becomes

$$\tilde{Y}_i = \sum_{j=1}^{p} \beta^{(j)} \tilde{X}_i^{(j)} + \text{noise}_i$$

which forces the regression surface through the center $(\tilde{x}^{(1)}, \ldots, \tilde{x}^{(p)}, \tilde{y}) = (0, 0, \ldots, 0)$ as with ordinary least squares. Note that it is not necessary to center the response variables when using the default offset value $\hat{f}^{[0]} = \overline{Y}$ in $L_2$Boosting. [For BinomialBoosting, we would center the predictor variables only but never the response, and we would use $\hat{f}^{[0]} \equiv \arg\min_c n^{-1} \sum_{i=1}^{n} \rho(Y_i, c)$.]

*Illustration*: *Prediction of total body fat.* Garcia et al. **(author?)** [34] report on the development of predictive regression equations for body fat content by means of $p = 9$ common anthropometric measurements which were obtained for $n = 71$ healthy German women. In addition, the women's body composition was measured by dual energy X-ray absorptiometry (DXA). This reference method is very accurate in measuring body fat but finds little applicability in practical environments, mainly because of high costs and the methodological efforts needed. Therefore, a simple regression equation for predicting DXA measurements of body fat is of special interest for the practitioner. Backward-elimination was applied to select important variables from the available anthropometrical measurements and Garcia et al. **(author?)** [34] report a final linear model utilizing hip circumference, knee breadth and a compound covariate which is defined as the sum of log chin skinfold, log triceps skinfold and log subscapular skinfold:

```
R> bf_lm <- lm(DEXfat ~ hipcirc
               + kneebreadth
               + anthro3a,
               data = bodyfat)
R> coef(bf_lm)
(Intercept) hipcirc kneebreadth anthro3a
  -75.23478 0.51153     1.90199  8.90964
```

A simple regression formula which is easy to communicate, such as a linear combination of only a few covariates, is of special interest in this application: we employ the `glmboost` function from package **mboost** to fit a linear regression model by means of $L_2$Boosting with componentwise linear least squares. By default, the function `glmboost` fits a linear model (with initial $m_{\text{stop}} = 100$ and shrinkage parameter $\nu = 0.1$) by minimizing squared error (argument `family = GaussReg()` is the default):

```
R> bf_glm <- glmboost(DEXfat ~ .,
              data = bodyfat,
              control= boost_control
              (center = TRUE))
```

Note that, by default, the mean of the response variable is used as an offset in the first step of the boosting algorithm. We center the covariates prior to model fitting in addition. As mentioned above, the special form of the base learner, that is, componentwise linear least squares, allows for a reformulation of the boosting fit in terms of a linear combination of the covariates which can be assessed via

```
R> coef(bf_glm)
  (Intercept)         age waistcirc  hipcirc
    0.000000    0.013602  0.189716 0.351626
elbowbreadth kneebreadth  anthro3a anthro3b
   -0.384140    1.736589  3.326860 3.656524
     anthro3c      anthro4
     0.595363     0.000000
attr(,"offset")
[1] 30.783
```

We notice that most covariates have been used for fitting and thus no extensive variable selection was performed in the above model. Thus, we need to investigate how many boosting iterations are appropriate. Resampling methods such as cross-validation or the bootstrap can be used to estimate the out-of-sample error for a varying number of boosting iterations. The out-of-bootstrap mean squared error for 100 bootstrap samples is depicted in the upper part of Figure 2. The plot leads to the impression that approximately $m_{\text{stop}} = 44$ would be a sufficient number of boosting iterations. In Section 5.4, a corrected version of the Akaike information criterion (AIC) is proposed for determining the optimal number of boosting iterations. This criterion attains its minimum for

```
R> mstop(aic <- AIC(bf_glm))
[1] 45
```

boosting iterations; see the bottom part of Figure 2 in addition. The coefficients of the linear model with $m_{\text{stop}} = 45$ boosting iterations are

```
R> coef(bf_glm[mstop(aic)])
  (Intercept)         age  waistcirc   hipcirc
   0.0000000   0.0023271  0.1893046 0.3488781
elbowbreadth kneebreadth   anthro3a  anthro3b
   0.0000000   1.5217686  3.3268603 3.6051548
     anthro3c      anthro4
    0.5043133    0.0000000
attr(,"offset")
[1] 30.783
```

and thus seven covariates have been selected for the final model (intercept equal to zero occurs here for mean centered response and predictors and hence, $n^{-1} \sum_{i=1}^{n} Y_i = 30.783$ is the intercept in the uncentered model). Note that the variables `hipcirc`, `kneebreadth` and `anthro3a`, which we have used for fitting a linear model at the beginning of this paragraph, have been selected by the boosting algorithm as well.

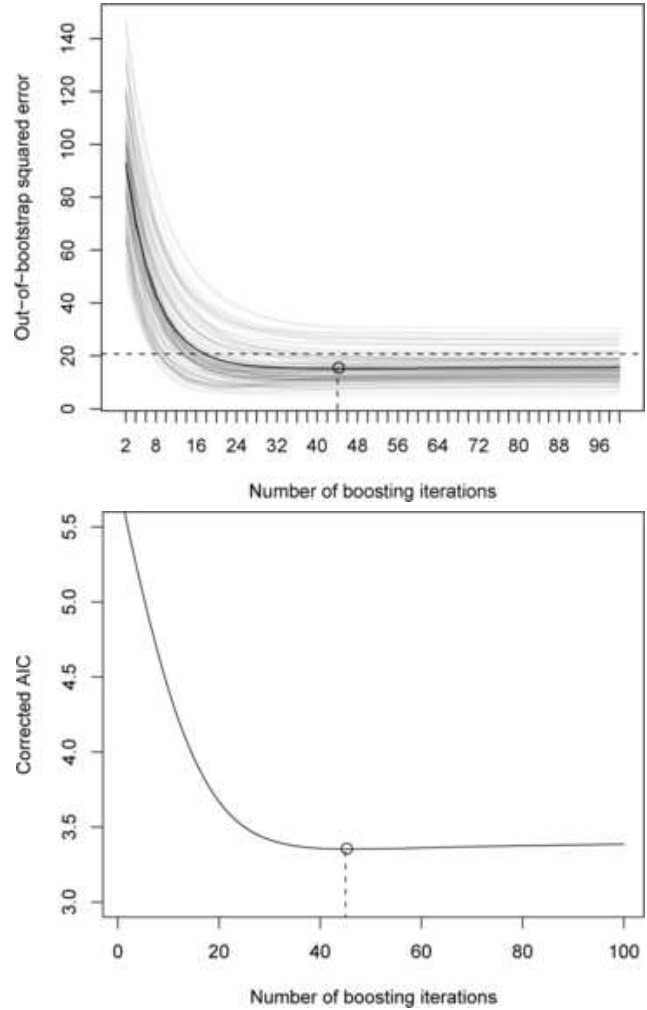

FIG. 2. *bodyfat* data: Out-of-bootstrap squared error for varying number of boosting iterations $m_{\text{stop}}$ (top). The dashed horizontal line depicts the average out-of-bootstrap error of the linear model for the preselected variables `hipcirc`, `kneebreadth` and `anthro3a` fitted via ordinary least squares. The lower part shows the corrected AIC criterion.

### 4.2 Componentwise Smoothing Spline for Additive Models

Additive and generalized additive models, introduced by Hastie and Tibshirani **(author?)** [40] (see also [41]), have become very popular for adding more flexibility to the linear structure in generalized linear models. Such flexibility can also be added in boosting (whose framework is especially useful for high-dimensional problems).

We can choose to use a nonparametric base procedure for function estimation. Suppose that

(4.2) $\hat{f}^{(j)}(\cdot)$ is a least squares cubic smoothing spline estimate based on $U_1, \ldots, U_n$ against $X_1^{(j)}, \ldots, X_n^{(j)}$ with fixed degrees of freedom df.

That is,

$$\hat{f}^{(j)}(\cdot) = \arg\min_{f(\cdot)} \sum_{i=1}^{n}(U_i - f(X_i^{(j)}))^2 \\ + \lambda \int (f''(x))^2\, dx, \quad (4.3)$$

where $\lambda > 0$ is a tuning parameter such that the trace of the corresponding hat matrix equals df. For further details, we refer to Green and Silverman **(author?)** [36]. As a note of caution, we use in the sequel the terminology of "hat matrix" in a broad sense: it is a linear operator but not a projection in general.

The base procedure is then

$$\hat{g}(x) = \hat{f}^{(\hat{\mathcal{S}})}(x^{(\hat{\mathcal{S}})}),$$
$$\hat{f}^{(j)}(\cdot) \text{ as above} \quad \text{and}$$
$$\hat{\mathcal{S}} = \arg\min_{1 \leq j \leq p} \sum_{i=1}^{n}(U_i - \hat{f}^{(j)}(X_i^{(j)}))^2,$$

where the degrees of freedom df are the same for all $\hat{f}^{(j)}(\cdot)$.

$L_2$Boosting with componentwise smoothing splines yields an additive model, including variable selection, that is, a fit which is additive in the predictor variables. This can be seen immediately since $L_2$Boosting proceeds additively for updating the function $\hat{f}^{[m]}(\cdot)$; see Section 3.3. We can normalize to obtain the following additive model estimator:

$$\hat{f}^{[m]}(x) = \hat{\mu} + \sum_{j=1}^{p} \hat{f}^{[m],(j)}(x^{(j)}),$$
$$n^{-1} \sum_{i=1}^{n} \hat{f}^{[m],(j)}(X_i^{(j)}) = 0 \quad \text{for all } j = 1, \ldots, p.$$

As with the componentwise linear least squares base procedure, we can use componentwise smoothing splines also in BinomialBoosting, yielding an additive logistic regression fit.

The degrees of freedom in the smoothing spline base procedure should be chosen "small" such as df = 4. This yields low variance but typically large bias of the base procedure. The bias can then be reduced by additional boosting iterations. This choice of low variance but high bias has been analyzed in Bühlmann and Yu **(author?)** [22]; see also Section 4.4.

Componentwise smoothing splines can be generalized to pairwise smoothing splines which search for and fit the best pairs of predictor variables such that smoothing of $U_1, \ldots, U_n$ against this pair of predictors reduces the residual sum of squares most. With $L_2$Boosting, this yields a nonparametric model fit with first-order interaction terms. The procedure has been empirically demonstrated to be often much better than fitting with MARS [23].

*Illustration*: *Prediction of total body fat (cont.)*. Being more flexible than the linear model which we fitted to the `bodyfat` data in Section 4.1, we estimate an additive model using the `gamboost` function from **mboost** (first with prespecified $m_{\text{stop}} = 100$ boosting iterations, $\nu = 0.1$ and squared error loss):

```
R> bf_gam
    <- gamboost(DEXfat ~ .,
                data = bodyfat)
```

The degrees of freedom in the componentwise smoothing spline base procedure can be defined by the `dfbase` argument, defaulting to 4.

We can estimate the number of boosting iterations $m_{\text{stop}}$ using the corrected AIC criterion described in Section 5.4 via

```
R> mstop(aic <- AIC(bf_gam))
[1] 46
```

Similarly to the linear regression model, the partial contributions of the covariates can be extracted from the boosting fit. For the most important variables, the partial fits are given in Figure 3 showing some slight nonlinearity, mainly for `kneebreadth`.

### 4.3 Trees

In the machine learning community, regression trees are the most popular base procedures. They have the advantage to be invariant under monotone transformations of predictor variables, that is, we do not need to search for good data transformations. Moreover, regression trees handle covariates measured at different scales (continuous, ordinal or nominal variables) in a unified way; unbiased split or variable selection in the context of different scales is proposed in [47].

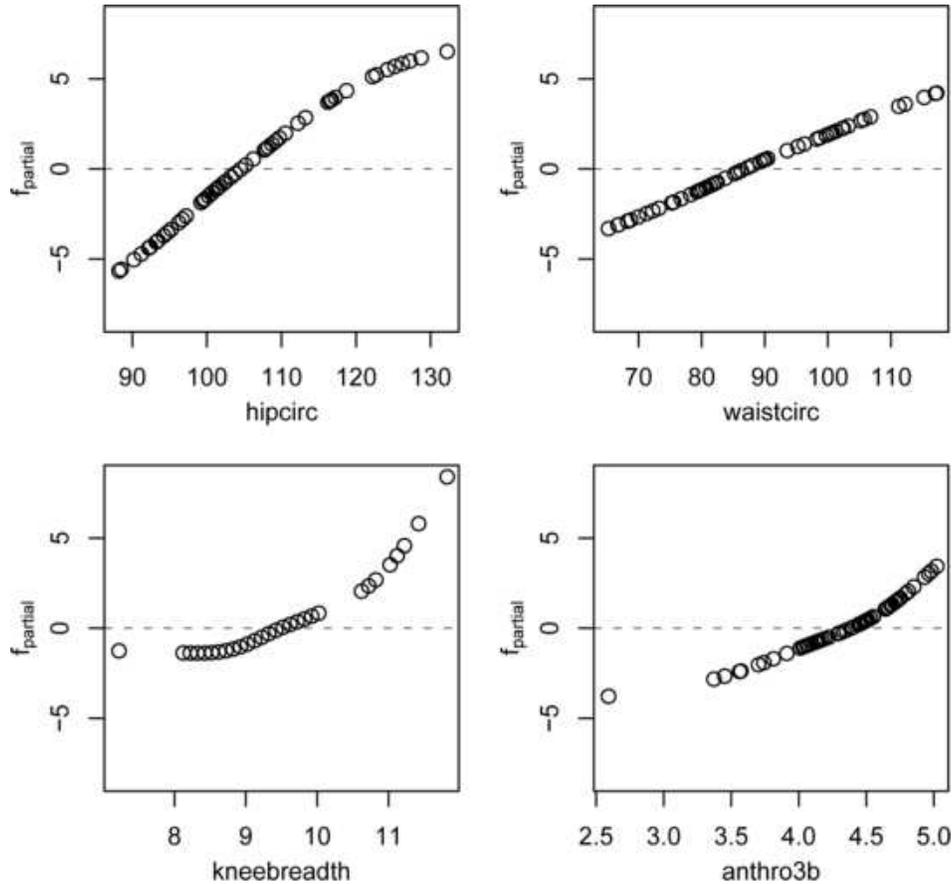

Fig. 3. `bodyfat` *data: Partial contributions of four covariates in an additive model (without centering of estimated functions to mean zero).*

When using stumps, that is, a tree with two terminal nodes only, the boosting estimate will be an additive model in the original predictor variables, because every stump-estimate is a function of a single predictor variable only. Similarly, boosting trees with (at most) $d$ terminal nodes result in a nonparametric model having at most interactions of order $d-2$. Therefore, if we want to constrain the degree of interactions, we can easily do this by constraining the (maximal) number of nodes in the base procedure.

*Illustration*: *Prediction of total body fat* (*cont.*). Both the **gbm** package [74] and the **mboost** package are helpful when decision trees are to be used as base procedures. In **mboost**, the function `blackboost` implements boosting for fitting such classical *black-box* models:

```
R> bf_black
   <- blackboost(DEXfat ~ .,
                 data = bodyfat,
                 control
                     = boost_control
                       (mstop = 500))
```

Conditional inference trees [47] as available from the **party** package [46] are utilized as base procedures. Here, the function `boost_control` defines the number of boosting iterations `mstop`.

Alternatively, we can use the function `gbm` from the **gbm** package which yields roughly the same fit as can be seen from Figure 4.

### 4.4 The Low-Variance Principle

We have seen above that the structural properties of a boosting estimate are determined by the choice of a base procedure. In our opinion, the structure specification should come first. After having made a choice, the question becomes how "complex" the base procedure should be. For example, how should we choose the degrees of freedom for the componentwise smoothing spline in (4.2)? A general answer is: choose the base procedure (having the desired structure) with low variance at the price of larger estimation bias. For the componentwise smoothing splines, this would imply a low number of degrees of freedom, for example, df = 4.

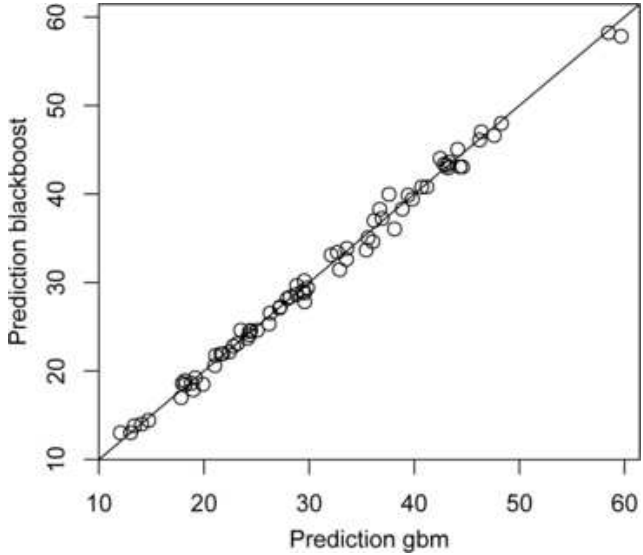

FIG. 4. *bodyfat data: Fitted values of both the* **gbm** *and* **mboost** *implementations of $L_2$Boosting with different regression trees as base learners.*

We give some reasons for the low-variance principle in Section 5.1 (Replica 1). Moreover, it has been demonstrated in Friedman **(author?)** [32] that a small step-size factor $\nu$ can be often beneficial and almost never yields substantially worse predictive performance of boosting estimates. Note that a small step-size factor can be seen as a shrinkage of the base procedure by the factor $\nu$, implying low variance but potentially large estimation bias.

## 5. $L_2$BOOSTING

$L_2$Boosting is functional gradient descent using the squared error loss which amounts to repeated fitting of ordinary residuals, as described already in Section 3.3.1. Here, we aim at increasing the understanding of the simple $L_2$Boosting algorithm. We first start with a toy problem of curve estimation, and we will then illustrate concepts and results which are especially useful for high-dimensional data. These can serve as heuristics for boosting algorithms with other convex loss functions for problems in for example, classification or survival analysis.

### 5.1 Nonparametric Curve Estimation: From Basics to Asymptotic Optimality

Consider the toy problem of estimating a regression function $\mathbb{E}[Y|X=x]$ with one-dimensional predictor $X \in \mathbb{R}$ and a continuous response $Y \in \mathbb{R}$.

Consider the case with a linear base procedure having a hat matrix $\mathcal{H}:\mathbb{R}^n \to \mathbb{R}^n$, mapping the response variables $\mathbf{Y} = (Y_1, \ldots, Y_n)^\top$ to their fitted values $(\hat{f}(X_1), \ldots, \hat{f}(X_n))^\top$. Examples include nonparametric kernel smoothers or smoothing splines. It is easy to show that the hat matrix of the $L_2$Boosting fit (for simplicity, with $\hat{f}^{[0]} \equiv 0$ and $\nu = 1$) in iteration $m$ equals

$$\begin{aligned}(5.1)\quad \mathcal{B}_m &= \mathcal{B}_{m-1} + \mathcal{H}(I - \mathcal{B}_{m-1})\\ &= I - (I - \mathcal{H})^m.\end{aligned}$$

Formula (5.1) allows for several insights. First, if the base procedure satisfies $\|I - \mathcal{H}\| < 1$ for a suitable norm, that is, has a "learning capacity" such that the residual vector is shorter than the input-response vector, we see that $\mathcal{B}_m$ converges to the identity $I$ as $m \to \infty$, and $\mathcal{B}_m \mathbf{Y}$ converges to the fully saturated model $\mathbf{Y}$, interpolating the response variables exactly. Thus, we see here explicitly that we have to stop early with the boosting iterations in order to prevent overfitting.

When specializing to the case of a cubic smoothing spline base procedure [cf. (4.3)], it is useful to invoke some eigenanalysis. The spectral representation is

$$\begin{aligned}\mathcal{H} &= UDU^\top,\\ U^\top U &= UU^\top = I,\\ D &= \operatorname{diag}(\lambda_1, \ldots, \lambda_n),\end{aligned}$$

where $\lambda_1 \geq \lambda_2 \geq \cdots \geq \lambda_n$ denote the (ordered) eigenvalues of $\mathcal{H}$. It then follows with (5.1) that

$$\begin{aligned}\mathcal{B}_m &= UD_m U^\top,\\ D_m &= \operatorname{diag}(d_{1,m}, \ldots, d_{n,m}),\\ d_{i,m} &= 1 - (1 - \lambda_i)^m.\end{aligned}$$

It is well known that a smoothing spline satisfies

$$\lambda_1 = \lambda_2 = 1, \quad 0 < \lambda_i < 1 \ (i = 3, \ldots, n).$$

Therefore, the eigenvalues of the boosting hat operator (matrix) in iteration $m$ satisfy

$$(5.2)\quad d_{1,m} \equiv d_{2,m} \equiv 1 \quad \text{for all } m,$$

$$(5.3)\quad \begin{aligned}0 < d_{i,m} &= 1 - (1 - \lambda_i)^m < 1 \quad (i = 3, \ldots, n),\\ d_{i,m} &\to 1 \quad (m \to \infty).\end{aligned}$$

When comparing the spectrum, that is, the set of eigenvalues, of a smoothing spline with its boosted version, we have the following. For both cases, the largest two eigenvalues are equal to 1. Moreover, all other eigenvalues can be changed either by varying the degrees of freedom df $= \sum_{i=1}^n \lambda_i$ in a single smoothing spline, or by varying the boosting iteration $m$ with some fixed (low-variance) smoothing spline base procedure having fixed (low) values $\lambda_i$. In Figure 5 we demonstrate the difference between the two approaches for changing "complexity" of the estimated curve fit by means of a toy example first shown in [22]. Both methods have about the same minimum mean squared error, but $L_2$Boosting overfits much more slowly than a single smoothing spline.

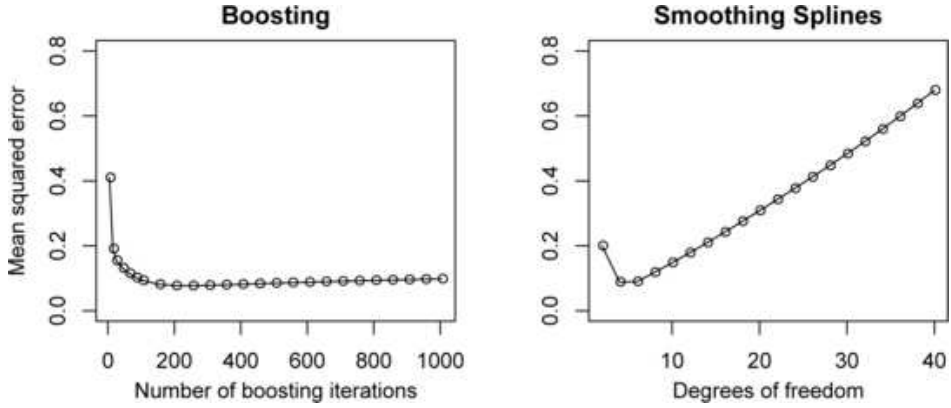

Fig. 5. *Mean squared prediction error* $\mathbb{E}[(f(X) - \hat{f}(X))^2]$ *for the regression model* $Y_i = 0.8X_i + \sin(6X_i) + \varepsilon_i$ $(i = 1,\ldots,n = 100)$, *with* $\varepsilon \sim \mathcal{N}(0,2), X_i \sim \mathcal{U}(-1/2,1/2)$, *averaged over* 100 *simulation runs. Left: $L_2$Boosting with smoothing spline base procedure (having fixed degrees of freedom df = 4) and using $\nu = 0.1$, for varying number of boosting iterations. Right: single smoothing spline with varying degrees of freedom.*

By careful inspection of the eigenanalysis for this simple case of boosting a smoothing spline, Bühlmann and Yu **(author?)** [22] proved an asymptotic minimax rate result:

REPLICA 1 ([22]). *When stopping the boosting iterations appropriately, that is, $m_{\text{stop}} = m_n = O(n^{4/(2\xi+1)})$, $m_n \to \infty$ $(n \to \infty)$ with $\xi \geq 2$ as below, $L_2$Boosting with cubic smoothing splines having fixed degrees of freedom achieves the minimax convergence rate over Sobolev function classes of smoothness degree $\xi \geq 2$, as $n \to \infty$.*

Two items are interesting. First, minimax rates are achieved by using a base procedure with fixed degrees of freedom which means low variance from an asymptotic perspective. Second, $L_2$Boosting with cubic smoothing splines has the capability to *adapt* to higher-order smoothness of the true underlying function; thus, with the stopping iteration as the one and only tuning parameter, we can nevertheless adapt to any higher-order degree of smoothness (without the need of choosing a higher-order spline base procedure).

Recently, asymptotic convergence and minimax rate results have been established for early-stopped boosting in more general settings [10, 91].

5.1.1 $L_2$*Boosting using kernel estimators.* As we have pointed out in Replica 1, $L_2$Boosting of smoothing splines can achieve faster mean squared error convergence rates than the classical $O(n^{-4/5})$, assuming that the true underlying function is sufficiently smooth. We illustrate here a related phenomenon with kernel estimators.

We consider fixed, univariate design points $x_i = i/n$ $(i = 1,\ldots,n)$ and the Nadaraya–Watson kernel estimator for the nonparametric regression function $\mathbb{E}[Y|X = x]$:

$$\hat{g}(x; h) = (nh)^{-1} \sum_{i=1}^n K\left(\frac{x - x_i}{h}\right) Y_i$$
$$= n^{-1} \sum_{i=1}^n K_h(x - x_i) Y_i,$$

where $h > 0$ is the bandwidth, $K(\cdot)$ is a kernel in the form of a probability density which is symmetric around zero and $K_h(x) = h^{-1}K(x/h)$. It is straightforward to derive the form of $L_2$Boosting using $m = 2$ iterations (with $\hat{f}^{[0]} \equiv 0$ and $\nu = 1$), that is, twicing [83], with the Nadaraya–Watson kernel estimator:

$$\hat{f}^{[2]}(x) = (nh)^{-1} \sum_{i=1}^n K_h^{\text{tw}}(x - x_i) Y_i,$$

$$K_h^{\text{tw}}(u) = 2K_h(u) - K_h * K_h(u),$$

where

$$K_h * K_h(u) = n^{-1} \sum_{r=1}^{n} K_h(u - x_r) K_h(x_r).$$

For fixed design points $x_i = i/n$, the kernel $K_h^{\text{tw}}(\cdot)$ is asymptotically equivalent to a higher-order kernel (which can take negative values) yielding a squared bias term of order $O(h^8)$, assuming that the true regression function is four times continuously differentiable. Thus, twicing or $L_2$Boosting with $m = 2$ iterations amounts to a Nadaraya–Watson kernel estimator with a higher-order kernel. This explains from another angle why boosting is able to improve the mean squared error rate of the base procedure. More details including also nonequispaced designs are given in DiMarzio and Taylor **(author?)** [27].

### 5.2 $L_2$Boosting for High-Dimensional Linear Models

Consider a potentially high-dimensional linear model

(5.4) $\quad Y_i = \beta_0 + \sum_{j=1}^{p} \beta^{(j)} X_i^{(j)} + \varepsilon_i, \quad i = 1, \ldots, n,$

where $\varepsilon_1, \ldots, \varepsilon_n$ are i.i.d. with $\mathbb{E}[\varepsilon_i] = 0$ and independent from all $X_i$'s. We allow for the number of predictors $p$ to be much larger than the sample size $n$. The model encompasses the representation of a noisy signal by an expansion with an overcomplete dictionary of functions $\{g^{(j)}(\cdot) : j = 1, \ldots, p\}$; for example, for surface modeling with design points in $Z_i \in \mathbb{R}^2$,

$$Y_i = f(Z_i) + \varepsilon_i,$$
$$f(z) = \sum_j \beta^{(j)} g^{(j)}(z) \quad (z \in \mathbb{R}^2).$$

Fitting the model (5.4) can be done using $L_2$Boosting with the componentwise linear least squares base procedure from Section 4.1 which fits in every iteration the best predictor variable reducing the residual sum of squares most. This method has the following basic properties:

1. As the number $m$ of boosting iterations increases, the $L_2$Boosting estimate $\hat{f}^{[m]}(\cdot)$ converges to a least squares solution. This solution is unique if the design matrix has full rank $p \leq n$.
2. When stopping early, which is usually needed to avoid overfitting, the $L_2$Boosting method often does variable selection.
3. The coefficient estimates $\hat{\beta}^{[m]}$ are (typically) shrunken versions of a least squares estimate $\hat{\beta}_{\text{OLS}}$, related to the Lasso as described in Section 5.2.1.

*Illustration*: *Breast cancer subtypes.* Variable selection is especially important in high-dimensional situations. As an example, we study a binary classification problem involving $p = 7129$ gene expression levels in $n = 49$ breast cancer tumor samples (data taken from [90]). For each sample, a binary response variable describes the lymph node status (25 negative and 24 positive).

The data are stored in form of an *exprSet* object westbc (see [35]) and we first extract the matrix of expression levels and the response variable:

```
R> x <- t(exprs(westbc))
R> y <- pData(westbc)$nodal.y
```

We aim at using $L_2$Boosting for classification (see Section 3.2.1), with classical AIC based on the binomial log-likelihood for stopping the boosting iterations. Thus, we first transform the factor y to a numeric variable with 0/1 coding:

```
R> yfit <- as.numeric(y) - 1
```

The general framework implemented in **mboost** allows us to specify the negative gradient (the ngradient argument) corresponding to the surrogate loss function, here the squared error loss implemented as a function rho, and a different evaluating loss function (the loss argument), here the negative binomial log-likelihood, with the Family function as follows:

```
R> rho <- function(y, f, w = 1) {
       p <- pmax(pmin(1 - 1e-05, f),
                 1e-05)
       -y * log(p) - (1 - y)
       * log(1 - p)
   }
R> ngradient
   <- function(y, f, w = 1) y - f
R> offset
   <- function(y, w)
       weighted.mean(y, w)
R> L2fm <- Family(ngradient =
                  ngradient,
                  loss = rho,
                  offset = offset)
```

The resulting object (called L2fm), bundling the negative gradient, the loss function and a function for computing an offset term (offset), can now be passed to the glmboost function for boosting with componentwise linear least squares (here initial $m_{\text{stop}} = 200$ iterations are used):

```
R> ctrl <- boost_control
              (mstop = 200,
               center = TRUE)
```

```
R> west_glm <- glmboost
                (x, yfit,
                 family = L2fm,
                 control = ctrl)
```

Fitting such a linear model to $p = 7129$ covariates for $n = 49$ observations takes about 3.6 seconds on a medium-scale desktop computer (Intel Pentium 4, 2.8 GHz). Thus, this form of estimation and variable selection is computationally very efficient. As a comparison, computing all Lasso solutions, using package **lars** [28, 39] in R (with use.Gram=FALSE), takes about 6.7 seconds.

The question how to choose $m_{\text{stop}}$ can be addressed by the classical AIC criterion as follows:

```
R> aic <- AIC(west_glm,
             method = "classical")
R> mstop(aic)
[1] 100
```

where the AIC is computed as $-2(\text{log-likelihood}) + 2(\text{degrees of freedom}) = 2(\text{evaluating loss}) + 2(\text{degrees of freedom})$; see (5.8). The notion of degrees of freedom is discussed in Section 5.3.

Figure 6 shows the AIC curve depending on the number of boosting iterations. When we stop after $m_{\text{stop}} = 100$ boosting iterations, we obtain 33 genes with nonzero regression coefficients whose standardized values $\hat{\beta}^{(j)}\sqrt{\widehat{\text{Var}}(X^{(j)})}$ are depicted in the left panel of Figure 6.

Of course, we could also use BinomialBoosting for analyzing the data; the computational CPU time would be of the same order of magnitude, that is, only a few seconds.

5.2.1 *Connections to the Lasso.* Hastie, Tibshirani and Friedman **(author?)** [42] pointed out first an intriguing connection between $L_2$Boosting with componentwise linear least squares and the Lasso [82] which is the following $\ell^1$-penalty method:

$$\hat{\beta}(\lambda) = \arg\min_{\beta} n^{-1} \sum_{i=1}^{n} \left(Y_i - \beta_0 - \sum_{j=1}^{p} \beta^{(j)} X_i^{(j)}\right)^2$$
(5.5)
$$+ \lambda \sum_{j=1}^{p} |\beta^{(j)}|.$$

Efron et al. **(author?)** [28] made the connection rigorous and explicit: they considered a version of $L_2$Boosting, called forward stagewise linear regression (FSLR), and they showed that FSLR with infinitesimally small step-sizes (i.e., the value $\nu$ in step 4 of the $L_2$Boosting algorithm in Section 3.3.1) produces a set of solutions which is approximately equivalent to the set of Lasso solutions when varying the regularization parameter $\lambda$ in Lasso [see (5.5)]. The approximate equivalence is derived by representing FSLR and Lasso as two different modifications of the computationally efficient least angle regression (LARS) algorithm from Efron et al. **(author?)** [28] (see also [68] for generalized linear models). The latter is very similar to the algorithm proposed earlier by Osborne, Presnell and Turlach **(author?)** [67]. In special cases where the design matrix satisfies a "positive cone condition," FSLR, Lasso and LARS all coincide ([28], page 425). For more general situations, when adding some backward steps to boosting, such modified $L_2$Boosting coincides with the Lasso (Zhao and Yu **(author?)** [93]).

Despite the fact that $L_2$Boosting and Lasso are not equivalent methods in general, it may be useful to interpret boosting as being "related" to $\ell^1$-penalty based methods.

5.2.2 *Asymptotic consistency in high dimensions.* We review here a result establishing asymptotic consistency for very high-dimensional but sparse linear models as in (5.4). To capture the notion of high-dimensionality, we equip the model with a dimensionality $p = p_n$ which is allowed to grow with sample size $n$; moreover, the coefficients $\beta^{(j)} = \beta_n^{(j)}$ are now potentially depending on $n$ and the regression function is denoted by $f_n(\cdot)$.

REPLICA 2 ([18]).   *Consider the linear model in* (5.4). *Assume that* $p_n = O(\exp(n^{1-\xi}))$ *for some* $0 < \xi \leq 1$ *(high-dimensionality) and* $\sup_{n \in \mathbb{N}} \sum_{j=1}^{p_n} |\beta_n^{(j)}| < \infty$ *(sparseness of the true regression function w.r.t. the* $\ell^1$*-norm); moreover, the variables* $X_i^{(j)}$ *are bounded and* $\mathbb{E}[|\varepsilon_i|^{4/\xi}] < \infty$. *Then: when stopping the boosting iterations appropriately, that is,* $m = m_n \to \infty$ $(n \to \infty)$ *sufficiently slowly,* $L_2$*Boosting with componentwise linear least squares satisfies*

$$\mathbb{E}_{X_{\text{new}}}[(\hat{f}_n^{[m_n]}(X_{\text{new}}) - f_n(X_{\text{new}}))^2] \to 0$$

*in probability* $(n \to \infty)$,

*where* $X_{\text{new}}$ *denotes new predictor variables, independent of and with the same distribution as the* $X$*-component of the data* $(X_i, Y_i)$ $(i = 1, \ldots, n)$.

The result holds for almost arbitrary designs and no assumptions about collinearity or correlations are required. Replica 2 identifies boosting as a method which is able to consistently estimate a very high-dimensional but sparse linear model; for the Lasso in (5.5), a similar result holds as well [37]. In terms of empirical performance, there seems to be no overall superiority of $L_2$Boosting over Lasso or vice versa.

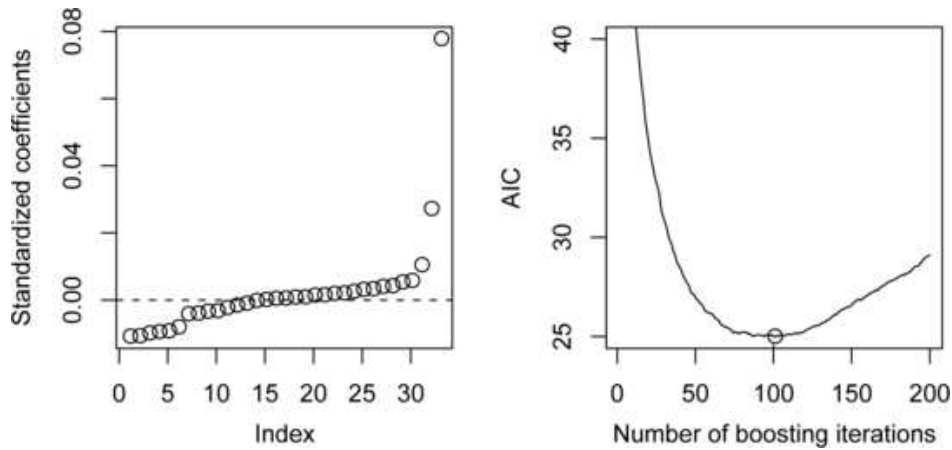

FIG. 6. `westbc` data: Standardized regression coefficients $\hat{\beta}^{(j)}\sqrt{\widehat{\mathrm{Var}}(X^{(j)})}$ (left panel) for $m_{\mathrm{stop}} = 100$ determined from the classical AIC criterion shown in the right panel.

5.2.3 *Transforming predictor variables.* In view of Replica 2, we may enrich the design matrix in model (5.4) with many transformed predictors: if the true regression function can be represented as a sparse linear combination of original or transformed predictors, consistency is still guaranteed. It should be noted, though, that the inclusion of noneffective variables in the design matrix does degrade the finite-sample performance to a certain extent.

For example, higher-order interactions can be specified in generalized AN(C)OVA models and $L_2$Boosting with componentwise linear least squares can be used to select a small number out of potentially many interaction terms.

As an option for continuously measured covariates, we may utilize a B-spline basis as illustrated in the next paragraph. We emphasize that during the process of $L_2$Boosting with componentwise linear least squares, individual spline basis functions from various predictor variables are selected and fitted one at a time; in contrast, $L_2$Boosting with componentwise smoothing splines fits a whole smoothing spline function (for a selected predictor variable) at a time.

*Illustration*: *Prediction of total body fat (cont.).* Such transformations and estimation of a corresponding linear model can be done with the `glmboost` function, where the model formula performs the computations of all transformations by means of the `bs` (B-spline basis) function from the package **splines**. First, we set up a formula transforming each covariate:

```
R> bsfm
DEXfat ~ bs(age) + bs(waistcirc) +
    bs(hipcirc) + bs(elbowbreadth) +
    bs(kneebreadth) + bs(anthro3a) +
    bs(anthro3b) + bs(anthro3c) +
    bs(anthro4)
```

and then fit the complex linear model by using the `glmboost` function with initial $m_{\mathrm{stop}} = 5000$ boosting iterations:

```
R> ctrl <- boost_control
            (mstop = 5000)
R> bf_bs <- glmboost
              (bsfm, data = bodyfat,
               control = ctrl)
R> mstop(aic <- AIC(bf_bs))
[1] 2891
```

The corrected AIC criterion (see Section 5.4) suggests to stop after $m_{\mathrm{stop}} = 2891$ boosting iterations and the final model selects 21 (transformed) predictor variables. Again, the partial contributions of each of the nine original covariates can be computed easily and are shown in Figure 7 (for the same variables as in Figure 3). Note that the depicted functional relationship derived from the model fitted above (Figure 7) is qualitatively the same as the one derived from the additive model (Figure 3).

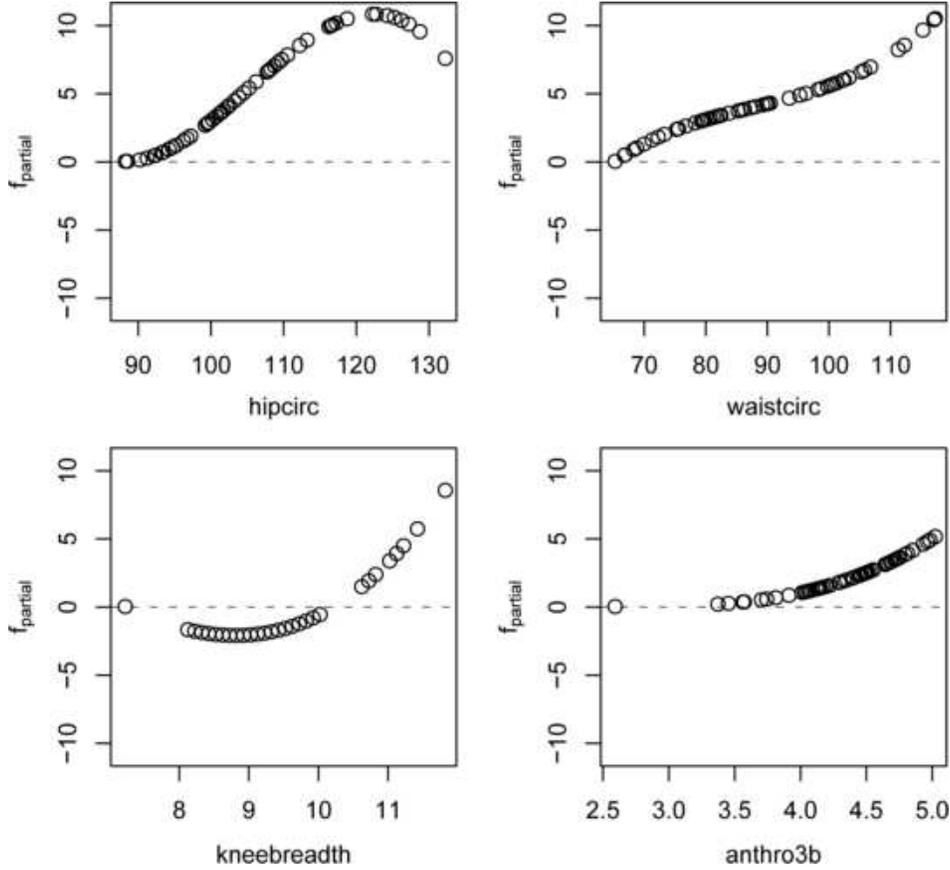

Fig. 7. *bodyfat* data: Partial fits for a linear model fitted to transformed covariates using B-splines (without centering of estimated functions to mean zero).

### 5.3 Degrees of Freedom for $L_2$Boosting

A notion of degrees of freedom will be useful for estimating the stopping iteration of boosting (Section 5.4).

5.3.1 *Componentwise linear least squares.* We consider $L_2$Boosting with componentwise linear least squares. Denote by

$$\mathcal{H}^{(j)} = \mathbf{X}^{(j)}(\mathbf{X}^{(j)})^\top / \|\mathbf{X}^{(j)}\|^2, \quad j = 1, \ldots, p,$$

the $n \times n$ hat matrix for the linear least squares fitting operator using the $j$th predictor variable $\mathbf{X}^{(j)} = (X_1^{(j)}, \ldots, X_n^{(j)})^\top$ only; $\|x\|^2 = x^\top x$ denotes the Euclidean norm for a vector $x \in \mathbb{R}^n$. The hat matrix of the componentwise linear least squares base procedure [see (4.1)] is then

$$\mathcal{H}^{(\hat{\mathcal{S}})} : (U_1, \ldots, U_n) \mapsto \hat{U}_1, \ldots, \hat{U}_n,$$

where $\hat{\mathcal{S}}$ is as in (4.1). Similarly to (5.1), we then obtain the hat matrix of $L_2$Boosting in iteration $m$:

$$\mathcal{B}_m = \mathcal{B}_{m-1} + \nu \cdot \mathcal{H}^{(\hat{\mathcal{S}}_m)}(I - \mathcal{B}_{m-1})$$

$$(5.6) \quad = I - (I - \nu\mathcal{H}^{(\hat{\mathcal{S}}_m)}) \cdot (I - \nu\mathcal{H}^{(\hat{\mathcal{S}}_{m-1})}) \cdots (I - \nu\mathcal{H}^{(\hat{\mathcal{S}}_1)}),$$

where $\hat{\mathcal{S}}_r \in \{1, \ldots, p\}$ denotes the component which is selected in the componentwise least squares base procedure in the $r$th boosting iteration. We emphasize that $\mathcal{B}_m$ is depending on the response variable $Y$ via the selected components $\hat{\mathcal{S}}_r$, $r = 1, \ldots, m$. Due to this dependence on $Y$, $\mathcal{B}_m$ should be viewed as an approximate hat matrix only. Neglecting the selection effect of $\hat{\mathcal{S}}_r$ ($r = 1, \ldots, m$), we define the degrees of freedom of the boosting fit in iteration $m$ as

$$\mathrm{df}(m) = \mathrm{trace}(\mathcal{B}_m).$$

Even with $\nu = 1$, $\mathrm{df}(m)$ is very different from counting the number of variables which have been selected until iteration $m$.

Having some notion of degrees of freedom at hand, we can estimate the error variance $\sigma_\varepsilon^2 = \mathbb{E}[\varepsilon_i^2]$ in the

linear model (5.4) by

$$\hat{\sigma}_\varepsilon^2 = \frac{1}{n - \mathrm{df}(m_{\mathrm{stop}})} \sum_{i=1}^n (Y_i - \hat{f}^{[m_{\mathrm{stop}}]}(X_i))^2.$$

Moreover, we can represent

(5.7) $$\mathcal{B}_m = \sum_{j=1}^p \mathcal{B}_m^{(j)},$$

where $\mathcal{B}_m^{(j)}$ is the (approximate) hat matrix which yields the fitted values for the $j$th predictor, that is, $\mathcal{B}_m^{(j)} \mathbf{Y} = \mathbf{X}^{(j)} \hat{\beta}_j^{[m]}$. Note that the $\mathcal{B}_m^{(j)}$'s can be easily computed in an iterative way by updating as follows:

$$\mathbf{B}_m^{(\hat{\mathcal{S}}_m)} = \mathcal{B}_{m-1}^{(\hat{\mathcal{S}}_m)} + \nu \cdot \mathcal{H}^{(\hat{\mathcal{S}}_m)}(I - \mathcal{B}_{m-1}),$$
$$\mathcal{B}_m^{(j)} = \mathcal{B}_{m-1}^{(j)} \quad \text{for all } j \neq \hat{\mathcal{S}}_m.$$

Thus, we have a decomposition of the total degrees of freedom into $p$ terms:

$$\mathrm{df}(m) = \sum_{j=1}^p \mathrm{df}^{(j)}(m),$$
$$\mathrm{df}^{(j)}(m) = \mathrm{trace}(\mathcal{B}_m^{(j)}).$$

The individual degrees of freedom $\mathrm{df}^{(j)}(m)$ are a useful measure to quantify the "complexity" of the individual coefficient estimate $\hat{\beta}_j^{[m]}$.

### 5.4 Internal Stopping Criteria for $L_2$Boosting

Having some degrees of freedom at hand, we can now use information criteria for estimating a good stopping iteration, without pursuing some sort of cross-validation.

We can use the corrected AIC [49]:

$$\mathrm{AIC}_c(m) = \log(\hat{\sigma}^2) + \frac{1 + \mathrm{df}(m)/n}{(1 - \mathrm{df}(m) + 2)/n},$$
$$\hat{\sigma}^2 = n^{-1} \sum_{i=1}^n (Y_i - (\mathcal{B}_m \mathbf{Y})_i)^2.$$

In **mboost**, the corrected AIC criterion can be computed via `AIC(x, method = "corrected")` (with x being an object returned by `glmboost` or `gamboost` called with `family = GaussReg()`). Alternatively, we may employ the gMDL criterion (Hansen and Yu **(author?)** [38]):

$$\mathrm{gMDL}(m) = \log(S) + \frac{\mathrm{df}(m)}{n} \log(F),$$
$$S = \frac{n\hat{\sigma}^2}{n - \mathrm{df}(m)}, F = \frac{\sum_{i=1}^n Y_i^2 - n\hat{\sigma}^2}{\mathrm{df}(m) S}.$$

The gMDL criterion bridges the AIC and BIC in a data-driven way: it is an attempt to adaptively select the better among the two.

When using $L_2$Boosting for binary classification (see also the end of Section 3.2 and the illustration in Section 5.2), we prefer to work with the binomial log-likelihood in AIC,

(5.8) $$\begin{aligned}\mathrm{AIC}(m) = &-2 \sum_{i=1}^n Y_i \log((\mathcal{B}_m \mathbf{Y})_i) \\ &+ (1 - Y_i) \log(1 - (\mathcal{B}_m \mathbf{Y})_i) \\ &+ 2 \mathrm{df}(m),\end{aligned}$$

or for BIC$(m)$ with the penalty term $\log(n)\mathrm{df}(m)$. (If $(\mathcal{B}_m \mathbf{Y})_i \notin [0,1]$, we truncate by $\max(\min((\mathcal{B}_m \mathbf{Y})_i, 1 - \delta), \delta)$ for some small $\delta > 0$, for example, $\delta = 10^{-5}$.)

## 6. BOOSTING FOR VARIABLE SELECTION

We address here the question whether boosting is a good variable selection scheme. For problems with many predictor variables, boosting is computationally much more efficient than classical all subset selection schemes. The mathematical properties of boosting for variable selection are still open questions, for example, whether it leads to a consistent model selection method.

### 6.1 $L_2$Boosting

When borrowing from the analogy of $L_2$Boosting with the Lasso (see Section 5.2.1), the following is relevant. Consider a linear model as in (5.4), allowing for $p \gg n$ but being sparse. Then, there is a sufficient and "almost" necessary neighborhood stability condition (the word "almost" refers to a strict inequality "<" whereas "≤" suffices for sufficiency) such that for some suitable penalty parameter $\lambda$ in (5.5), the Lasso finds the true underlying submodel (the predictor variables with corresponding regression coefficients $\neq 0$) with probability tending quickly to 1 as $n \to \infty$ [65]. It is important to note the role of the sufficient and "almost" necessary condition of the Lasso for model selection: Zhao and Yu **(author?)** [94] call it the "irrepresentable condition" which has (mainly) implications on the "degree of collinearity" of the design (predictor variables), and they give examples where it holds and where it fails to be true. A further complication is the fact that when tuning the Lasso for prediction optimality, that is, choosing the penalty parameter $\lambda$ in (5.5) such that the mean squared error is minimal, the probability for estimating the true submodel converges to a number which is less than 1 or even zero if the problem is high-dimensional [65]. In fact, the prediction optimal tuned Lasso selects asymptotically too large models.

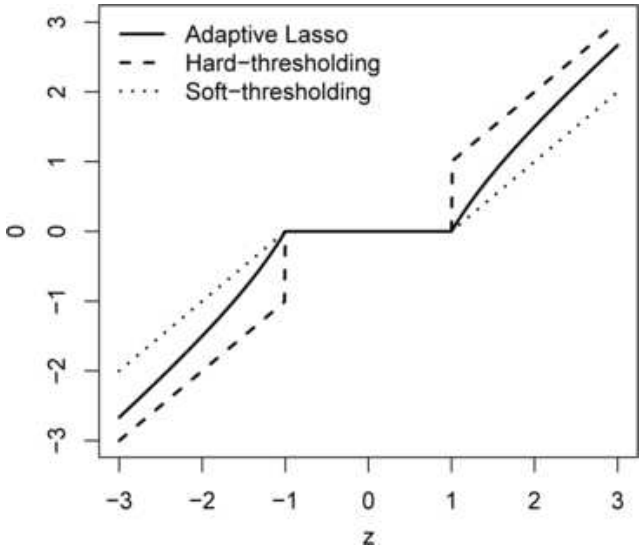

FIG. 8. *Hard-threshold (dotted-dashed), soft-threshold (dotted) and adaptive Lasso (solid) estimator in a linear model with orthonormal design. For this design, the adaptive Lasso coincides with the nonnegative garrote* [13]. *The value on the x-abscissa, denoted by z, is a single component of* $\mathbf{X}^\top \mathbf{Y}$.

The bias of the Lasso mainly causes the difficulties mentioned above. We often would like to construct estimators which are less biased. It is instructive to look at regression with orthonormal design, that is, the model (5.4) with $\sum_{i=1}^n X_i^{(j)} X_i^{(k)} = \delta_{jk}$. Then, the Lasso and also $L_2$Boosting with componentwise linear least squares and using very small $\nu$ (in step 4 of $L_2$Boosting; see Section 3.3.1) yield the soft-threshold estimator [23, 28]; see Figure 8. It exhibits the same amount of bias regardless by how much the observation (the variable $z$ in Figure 8) exceeds the threshold. This is in contrast to the hard-threshold estimator and the adaptive Lasso in (6.1) which are much better in terms of bias.

Nevertheless, the (computationally efficient) Lasso seems to be a very useful method for variable filtering: for many cases, the prediction optimal tuned Lasso selects a submodel which contains the true model with high probability. A nice proposal to correct Lasso's overestimation behavior is the adaptive Lasso, given by Zou **(author?)** [96]. It is based on reweighting the penalty function. Instead of (5.5), the adaptive Lasso estimator is

$$\hat{\beta}(\lambda) = \arg\min_\beta n^{-1} \sum_{i=1}^n \left( Y_i - \beta_0 - \sum_{j=1}^p \beta^{(j)} X_i^{(j)} \right)^2$$
(6.1)
$$+ \lambda \sum_{j=1}^p \frac{|\beta^{(j)}|}{|\hat{\beta}_{\text{init}}^{(j)}|},$$

where $\hat{\beta}_{\text{init}}$ is an initial estimator, for example, the Lasso (from a first stage of Lasso estimation). Consistency of the adaptive Lasso for variable selection has been proved for the case with fixed predictor-dimension $p$ [96] and also for the high-dimensional case with $p = p_n \gg n$ [48].

We do not expect that boosting is free from the difficulties which occur when using the Lasso for variable selection. The hope is, though, that also boosting would produce an interesting set of submodels when varying the number of iterations.

### 6.2 Twin Boosting

Twin Boosting [19] is the boosting analogue to the adaptive Lasso. It consists of two stages of boosting: the first stage is as usual, and the second stage is enforced to resemble the first boosting round. For example, if a variable has not been selected in the first round of boosting, it will not be selected in the second; this property also holds for the adaptive Lasso in (6.1), that is, $\hat{\beta}_{\text{init}}^{(j)} = 0$ enforces $\hat{\beta}^{(j)} = 0$. Moreover, Twin Boosting with componentwise linear least squares is proved to be equivalent to the adaptive Lasso for the case of an orthonormal linear model and it is empirically shown, in general and for various base procedures and models, that it has much better variable selection properties than the corresponding boosting algorithm [19]. In special settings, similar results can be obtained with Sparse Boosting [23]; however, Twin Boosting is much more generically applicable.

## 7. BOOSTING FOR EXPONENTIAL FAMILY MODELS

For exponential family models with general loss functions, we can use the generic FGD algorithm as described in Section 2.1.

First, we address the issue about omitting a line search between steps 3 and 4 of the generic FGD algorithm. Consider the empirical risk at iteration

$m$,

$$n^{-1} \sum_{i=1}^{n} \rho(Y_i, \hat{f}^{[m]}(X_i))$$

$$(7.1) \quad \approx n^{-1} \sum_{i=1}^{n} \rho(Y_i, \hat{f}^{[m-1]}(X_i))$$

$$- \nu n^{-1} \sum_{i=1}^{n} U_i \hat{g}^{[m]}(X_i),$$

using a first-order Taylor expansion and the definition of $U_i$. Consider the case with the componentwise linear least squares base procedure and without loss of generality with standardized predictor variables [i.e., $n^{-1} \sum_{i=1}^{n} (X_i^{(j)})^2 = 1$ for all $j$]. Then,

$$\hat{g}^{[m]}(x) = n^{-1} \sum_{i=1}^{n} U_i X_i^{(\hat{\mathcal{S}}_m)} x^{(\hat{\mathcal{S}}_m)},$$

and the expression in (7.1) becomes

$$n^{-1} \sum_{i=1}^{n} \rho(Y_i, \hat{f}^{[m]}(X_i))$$

$$(7.2) \quad \approx n^{-1} \sum_{i=1}^{n} \rho(Y_i, \hat{f}^{[m-1]}(X_i))$$

$$- \nu \left(n^{-1} \sum_{i=1}^{n} U_i X_i^{(\hat{\mathcal{S}}_m)}\right)^2.$$

In case of the squared error loss $\rho_{L_2}(y, f) = |y - f|^2/2$, we obtain the exact identity:

$$n^{-1} \sum_{i=1}^{n} \rho_{L_2}(Y_i, \hat{f}^{[m]}(X_i))$$

$$= n^{-1} \sum_{i=1}^{n} \rho_{L_2}(Y_i, \hat{f}^{[m-1]}(X_i))$$

$$- \nu(1 - \nu/2) \left(n^{-1} \sum_{i=1}^{n} U_i X_i^{(\hat{\mathcal{S}}_m)}\right)^2.$$

Comparing this with (7.2), we see that functional gradient descent with a general loss function and without additional line-search behaves very similarly to $L_2$Boosting (since $\nu$ is small) with respect to optimizing the empirical risk; for $L_2$Boosting, the numerical convergence rate is $n^{-1} \sum_{i=1}^{n} \rho_{L_2}(Y_i, \hat{f}^{[m]}(X_i)) = O(m^{-1/6})$ $(m \to \infty)$ [81]. This completes our reasoning why the line-search in the general functional gradient descent algorithm can be omitted, of course at the price of doing more iterations but not necessarily more computing time (since the line-search is omitted in every iteration).

### 7.1 BinomialBoosting

For binary classification with $Y \in \{0, 1\}$, BinomialBoosting uses the negative binomial log-likelihood from (3.1) as loss function. The algorithm is described in Section 3.3.2. Since the population minimizer is $f^*(x) = \log[p(x)/(1 - p(x))]/2$, estimates from BinomialBoosting are on half of the logit-scale: the componentwise linear least squares base procedure yields a logistic linear model fit while using componentwise smoothing splines fits a logistic additive model. Many of the concepts and facts from Section 5 about $L_2$Boosting become useful heuristics for BinomialBoosting.

One principal difference is the derivation of the boosting hat matrix. Instead of (5.6), a linearization argument leads to the following recursion [assuming $\hat{f}^{[0]}(\cdot) \equiv 0$] for an approximate hat matrix $\mathcal{B}_m$:

$$\mathcal{B}_1 = \nu 4 W^{[0]} \mathcal{H}^{(\hat{\mathcal{S}}_1)},$$

$$(7.3) \quad \mathcal{B}_m = \mathcal{B}_{m-1} + 4\nu W^{[m-1]} \mathcal{H}^{(\hat{\mathcal{S}}_m)} (I - \mathcal{B}_{m-1})$$

$$(m \geq 2),$$

$$W^{[m]} = \text{diag}(\hat{p}^{[m]}(X_i)(1 - \hat{p}^{[m]}(X_i); \ 1 \leq i \leq n)).$$

A derivation is given in Appendix A.2. Degrees of freedom are then defined as in Section 5.3,

$$\text{df}(m) = \text{trace}(\mathcal{B}_m),$$

and they can be used for information criteria, for example,

$$\text{AIC}(m) = -2 \sum_{i=1}^{n} [Y_i \log(\hat{p}^{[m]}(X_i))$$

$$+ (1 - Y_i) \log(1 - \hat{p}^{[m]}(X_i))]$$

$$+ 2 \text{df}(m),$$

or for BIC$(m)$ with the penalty term $\log(n) \text{df}(m)$. In **mboost**, this AIC criterion can be computed via `AIC(x, method = "classical")` (with x being an object returned by `glmboost` or `gamboost` called with `family = Binomial()`).

*Illustration*: *Wisconsin prognostic breast cancer*. Prediction models for recurrence events in breast cancer patients based on covariates which have been computed from a digitized image of a fine needle aspirate of breast tissue (those measurements describe characteristics of the cell nuclei present in the image) have been studied by Street, Mangasarian and Wolberg **(author?)** [80] (the data are part of the UCI repository [11]).

We first analyze these data as a binary prediction problem (recurrence vs. nonrecurrence) and later in Section 8 by means of survival models. We are faced with many covariates ($p = 32$) for a limited number of observations without missing values ($n = 194$), and variable selection is an important issue. We can choose a classical logistic regression model via AIC in a stepwise algorithm as follows:

```
R> cc <- complete.cases(wpbc)
R> wpbc2
     <- wpbc[cc,
             colnames(wpbc) != "time"]
R> wpbc_step
     <- step(glm(status ~ .,
                 data = wpbc2,
                 family = binomial()),
             trace = 0)
```

The final model consists of 16 parameters with
```
R> logLik(wpbc_step)
'log Lik.' -80.13 (df=16)
R> AIC(wpbc_step)
[1] 192.26
```

and we want to compare this model to a logistic regression model fitted via gradient boosting. We simply select the `Binomial` family [with default offset of $1/2 \log(\hat{p}/(1-\hat{p}))$, where $\hat{p}$ is the empirical proportion of recurrences] and we initially use $m_{\text{stop}} = 500$ boosting iterations:

```
R> ctrl <- boost_control
              (mstop = 500,
               center = TRUE)
R> wpbc_glm
     <- glmboost(status ~ .,
                 data = wpbc2,
                 family = Binomial(),
                 control = ctrl)
```

The classical AIC criterion ($-2\log$-likelihood$+2$df) suggests to stop after
```
R> aic <- AIC(wpbc_glm, "classical")
R> aic
[1] 199.54
Optimal number of boosting iterations: 465
Degrees of freedom (for mstop = 465): 9.147
```
boosting iterations. We now restrict the number of boosting iterations to $m_{\text{stop}} = 465$ and then obtain the estimated coefficients via
```
R> wpbc_glm <- wpbc_glm[mstop(aic)]
R> coef(wpbc_glm)
       [abs(coef(wpbc_glm)) > 0]
```

```
   (Intercept)       mean_radius      mean_texture
    -1.2511e-01       -5.8453e-03      -2.4505e-02
mean_smoothness      mean_symmetry   mean_fractaldim
     2.8513e+00       -3.9307e+00      -2.8253e+01
    SE_texture        SE_perimeter    SE_compactness
    -8.7553e-02        5.4917e-02       1.1463e+01
   SE_concavity     SE_concavepoints    SE_symmetry
    -6.9238e+00       -2.0454e+01       5.2125e+00
  SE_fractaldim      worst_radius    worst_perimeter
     5.2187e+00        1.3468e-02       1.2108e-03
    worst_area    worst_smoothness  worst_compactness
     1.8646e-04        9.9560e+00      -1.9469e-01
         tsize            pnodes
     4.1561e-02        2.4445e-02
```

(Because of using the offset-value $\hat{f}^{[0]}$, we have to add the value $\hat{f}^{[0]}$ to the reported intercept estimate above for the logistic regression model.)

A generalized additive model adds more flexibility to the regression function but is still interpretable. We fit a logistic additive model to the `wpbc` data as follows:

```
R> wpbc_gam <- gamboost(status ~ .,
          data = wpbc2,
          family = Binomial())
R> mopt <- mstop(aic <-
          AIC(wpbc_gam, "classical"))
R> aic
[1] 199.76
Optimal number of boosting iterations: 99
Degrees of freedom (for mstop = 99): 14.583
```

This model selected 16 out of 32 covariates. The partial contributions of the four most important variables are depicted in Figure 9 indicating a remarkable degree of nonlinearity.

### 7.2 PoissonBoosting

For count data with $Y \in \{0, 1, 2, \ldots\}$, we can use Poisson regression: we assume that $Y|X = x$ has a Poisson($\lambda(x)$) distribution and the goal is to estimate the function $f(x) = \log(\lambda(x))$. The negative log-likelihood yields then the loss function

$$\rho(y, f) = -yf + \exp(f), \quad f = \log(\lambda),$$

which can be used in the functional gradient descent algorithm in Section 2.1, and it is implemented in **mboost** as `Poisson()` family.

Similarly to (7.3), the approximate boosting hat matrix is computed by the following recursion:

$$\mathcal{B}_1 = \nu W^{[0]} \mathcal{H}^{(\hat{\mathcal{S}}_1)},$$

(7.4) $\quad \mathcal{B}_m = \mathcal{B}_{m-1} + \nu W^{[m-1]} \mathcal{H}^{(\hat{\mathcal{S}}_m)}(I - \mathcal{B}_{m-1})$

$$(m \geq 2),$$

$$W^{[m]} = \text{diag}(\hat{\lambda}^{[m]}(X_i); \ 1 \leq i \leq n).$$

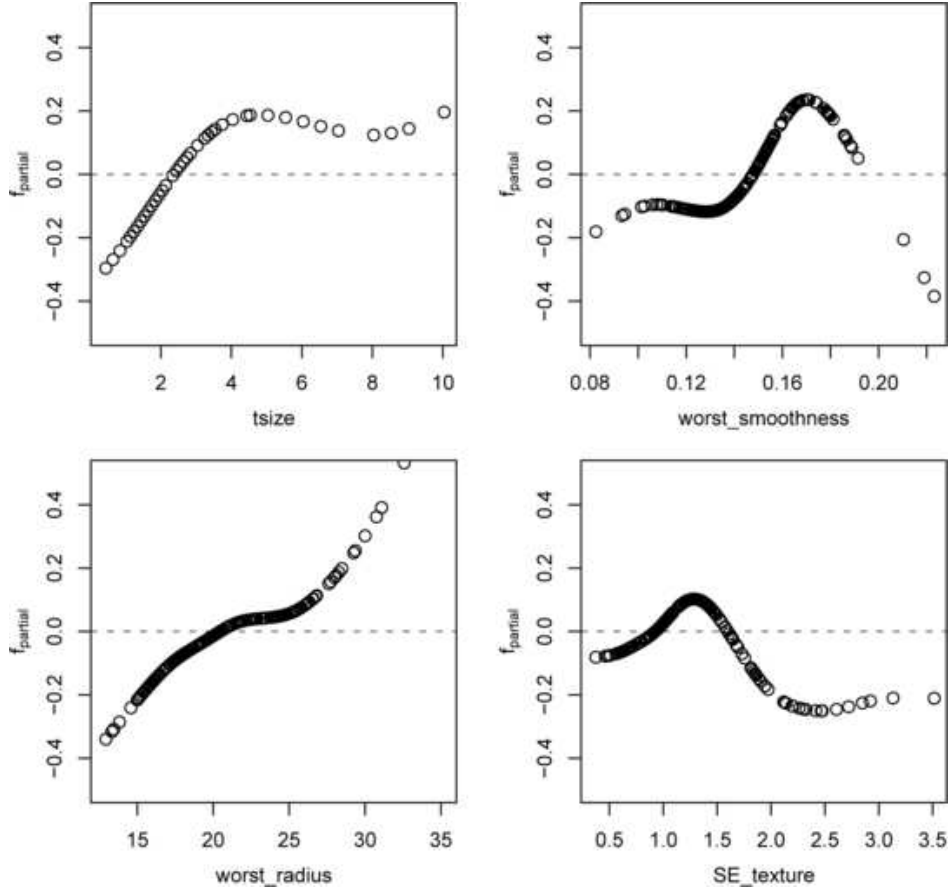

Fig. 9. *wpbc* data: Partial contributions of four selected covariates in an additive logistic model (without centering of estimated functions to mean zero).

### 7.3 Initialization of Boosting

We have briefly described in Sections 2.1 and 4.1 the issue of choosing an initial value $\hat{f}^{[0]}(\cdot)$ for boosting. This can be quite important for applications where we would like to estimate some parts of a model in an unpenalized (nonregularized) fashion, with others being subject to regularization.

For example, we may think of a parametric form of $\hat{f}^{[0]}(\cdot)$, estimated by maximum likelihood, and deviations from the parametric model would be built in by pursuing boosting iterations (with a nonparametric base procedure). A concrete example would be: $\hat{f}^{[0]}(\cdot)$ is the maximum likelihood estimate in a generalized linear model and boosting would be done with componentwise smoothing splines to model additive deviations from a generalized linear model. A related strategy has been used in [4] for modeling multivariate volatility in financial time series.

Another example would be a linear model $\mathbf{Y} = \mathbf{X}\beta + \varepsilon$ as in (5.4) where some of the predictor variables, say the first $q$ predictor variables $X^{(1)}, \ldots, X^{(q)}$, enter the estimated linear model in an unpenalized way. We propose to do ordinary least squares regression on $X^{(1)}, \ldots, X^{(q)}$: consider the projection $P_q$ onto the linear span of $X^{(1)}, \ldots, X^{(q)}$ and use $L_2$Boosting with componentwise linear least squares on the new response $(I - P_q)\mathbf{Y}$ and the new $(p - q)$-dimensional predictor $(I - P_q)\mathbf{X}$. The final model estimate is then $\sum_{j=1}^{q} \hat{\beta}_{\text{OLS},j} x^{(j)} + \sum_{j=q+1}^{p} \hat{\beta}_j^{[m_{\text{stop}}]} \tilde{x}^{(j)}$, where the latter part is from $L_2$Boosting and $\tilde{x}^{(j)}$ is the residual when linearly regressing $x^{(j)}$ to $x^{(1)}, \ldots, x^{(q)}$. A special case which is used in most applications is with $q = 1$ and $X^{(1)} \equiv 1$ encoding for an intercept. Then, $(I - P_1)\mathbf{Y} = \mathbf{Y} - \overline{Y}$ and $(I - P_1)\mathbf{X}^{(j)} = \mathbf{X}^{(j)} - n^{-1} \sum_{i=1}^{n} X_i^{(j)}$. This is exactly the proposal at the end of Section 4.1. For generalized linear models, analogous concepts can be used.

## 8. SURVIVAL ANALYSIS

The negative gradient of Cox's partial likelihood can be used to fit proportional hazards models to censored response variables with boosting algorithms [71]. Of course, all types of base procedures can be utilized; for example, componentwise linear least squares fits a Cox model with a linear predictor.

Alternatively, we can use the weighted least squares framework with weights arising from inverse probability censoring. We sketch this approach in the sequel; details are given in [45]. We assume complete data of the following form: survival times $T_i \in \mathbb{R}^+$ (some of them right-censored) and predictors $X_i \in \mathbb{R}^p$, $i = 1, \ldots, n$. We transform the survival times to the log-scale, but this step is not crucial for what follows: $Y_i = \log(T_i)$. What we observe is

$$O_i = (\tilde{Y}_i, X_i, \Delta_i),$$
$$\tilde{Y}_i = \log(\tilde{T}_i),$$
$$\tilde{T}_i = \min(T_i, C_i),$$

where $\Delta_i = I(T_i \leq C_i)$ is a censoring indicator and $C_i$ is the censoring time. Here, we make a restrictive assumption that $C_i$ is conditionally independent of $T_i$ given $X_i$ (and we assume independence among different indices $i$); this implies that the coarsening at random assumption holds [89].

We consider the squared error loss for the complete data, $\rho(y, f) = |y - f|^2$ (without the irrelevant factor 1/2). For the observed data, the following weighted version turns out to be useful:

$$\rho_{\text{obs}}(o, f) = (\tilde{y} - f)^2 \Delta \frac{1}{G(\tilde{t}|x)},$$
$$G(c|x) = \mathbb{P}[C > c|X = x].$$

Thus, the observed data loss function is weighted by the inverse probability for censoring $\Delta G(\tilde{t}|x)^{-1}$ (the weights are inverse probabilities of censoring; IPC). Under the coarsening at random assumption, it then holds that

$$\mathbb{E}_{Y,X}[(Y - f(X))^2] = \mathbb{E}_O[\rho_{\text{obs}}(O, f(X))];$$

see van der Laan and Robins **(author?)** [89].

The strategy is then to estimate $G(\cdot|x)$, for example, by the Kaplan–Meier estimator, and do weighted $L_2$Boosting using the weighted squared error loss:

$$\sum_{i=1}^n \Delta_i \frac{1}{\hat{G}(\tilde{T}_i|X_i)}(\tilde{Y}_i - f(X_i))^2,$$

where the weights are of the form $\Delta_i \hat{G}(\tilde{T}_i|X_i)^{-1}$ (the specification of the estimator $\hat{G}(t|x)$ may play a substantial role in the whole procedure). As demonstrated in the previous sections, we can use various base procedures as long as they allow for weighted least squares fitting. Furthermore, the concepts of degrees of freedom and information criteria are analogous to Sections 5.3 and 5.4. Details are given in [45].

*Illustration*: *Wisconsin prognostic breast cancer* (*cont.*). Instead of the binary response variable describing the recurrence status, we make use of the additionally available time information for modeling the time to recurrence; that is, all observations with nonrecurrence are censored. First, we calculate IPC weights:

```
R> censored <- wpbc$status == "R"
R> iw <- IPCweights(Surv(wpbc$time,
                         censored))
R> wpbc3 <- wpbc[, names(wpbc) !=
                "status"]
```

and fit a weighted linear model by boosting with componentwise linear weighted least squares as base procedure:

```
R> ctrl <- boost_control(
       mstop = 500, center = TRUE)
R> wpbc_surv <- glmboost(
       log(time) ~ ., data = wpbc3,
       control = ctrl, weights = iw)
R> mstop(aic <- AIC(wpbc_surv))
[1] 122
R> wpbc_surv <- wpbc_surv[
       mstop(aic)]
```

The following variables have been selected for fitting:

```
R> names(coef(wpbc_surv)
           [abs(coef(wpbc_surv)) > 0])
[1] "mean_radius"     "mean_texture"
[3] "mean_perimeter"  "mean_smoothness"
[5] "mean_symmetry"   "SE_texture"
[7] "SE_smoothness"   "SE_concavepoints"
[9] "SE_symmetry"     "worst_concavepoints"
```

and the fitted values are depicted in Figure 10, showing a reasonable model fit.

Alternatively, a Cox model with linear predictor can be fitted using $L_2$Boosting by implementing the negative gradient of the partial likelihood (see [71]) via

```
R> ctrl <- boost_control
              (center = TRUE)
R> glmboost
       (Surv(wpbc$time,
             wpbc$status == "N") ~ .,
        data = wpbc,
        family = CoxPH(),
        control = ctrl)
```

For more examples, such as fitting an additive Cox model using **mboost**, see [44].

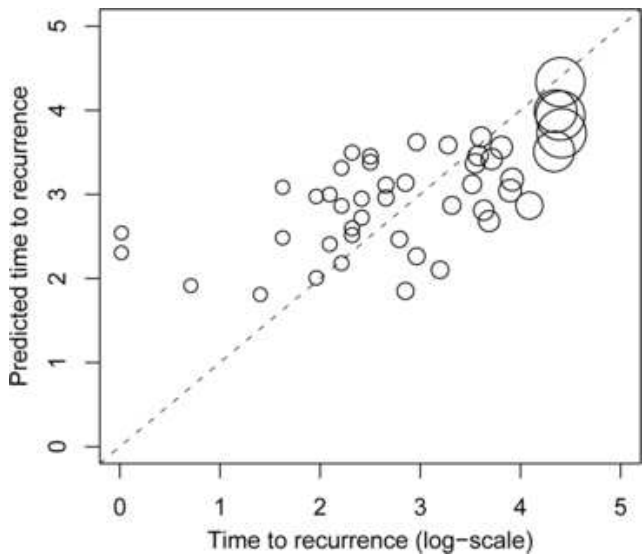

FIG. 10. `wpbc` *data: Fitted values of an IPC-weighted linear model, taking both time to recurrence and censoring information into account. The radius of the circles is proportional to the IPC weight of the corresponding observation; censored observations with IPC weight zero are not plotted.*

## 9. OTHER WORKS

We briefly summarize here some other works which have not been mentioned in the earlier sections. A very different exposition than ours is the overview of boosting by Meir and Rätsch **(author?)** [66].

### 9.1 Methodology and Applications

Boosting methodology has been used for various other statistical models than what we have discussed in the previous sections. Models for multivariate responses are studied in [20, 59]; some multiclass boosting methods are discussed in [33, 95]. Other works deal with boosting approaches for generalized linear and nonparametric models [55, 56, 85, 86], for flexible semiparametric mixed models [88] or for nonparametric models with quality constraints [54, 87]. Boosting methods for estimating propensity scores, a special weighting scheme for modeling observational data, are proposed in [63].

There are numerous applications of boosting methods to real data problems. We mention here classification of tumor types from gene expressions [25, 26], multivariate financial time series [2, 3, 4], text classification [78], document routing [50] or survival analysis [8] (different from the approach in Section 8).

### 9.2 Asymptotic Theory

The asymptotic analysis of boosting algorithms includes consistency and minimax rate results. The first consistency result for AdaBoost has been given by Jiang **(author?)** [51], and a different constructive proof with a range for the stopping value $m_{\text{stop}} = m_{\text{stop},n}$ is given in [7]. Later, Zhang and Yu **(author?)** [92] generalized the results for a functional gradient descent with an additional relaxation scheme, and their theory covers also more general loss functions than the exponential loss in AdaBoost. For $L_2$Boosting, the first minimax rate result has been established by Bühlmann and Yu **(author?)** [22]. This has been extended to much more general settings by Yao, Rosasco and Caponnetto **(author?)** [91] and Bissantz et al. **(author?)** [10].

In the machine learning community, there has been a substantial focus on estimation in the convex hull of function classes (cf. [5, 6, 58]). For example, one may want to estimate a regression or probability function by using

$$\sum_{k=1}^{\infty} \hat{w}_k \hat{g}^{[k]}(\cdot), \quad \hat{w}_k \geq 0, \quad \sum_{k=1}^{\infty} \hat{w}_k = 1,$$

where the $\hat{g}^{[k]}(\cdot)$'s belong to a function class such as stumps or trees with a fixed number of terminal nodes. The estimator above is a convex combination of individual functions, in contrast to boosting which pursues a linear combination. By scaling, which is necessary in practice and theory (cf. [58]), one can actually look at this as a linear combination of functions whose coefficients satisfy $\sum_k \hat{w}_k = \lambda$. This then represents an $\ell^1$-constraint as in Lasso, a relation which we have already seen from another perspective in Section 5.2.1. Consistency of such convex combination or $\ell^1$-regularized "boosting" methods has been given by Lugosi and Vayatis **(author?)** [58]. Mannor, Meir and Zhang **(author?)** [61] and Blanchard, Lugosi and Vayatis **(author?)** [12] derived results for rates of convergence of (versions of) convex combination schemes.

## APPENDIX A.1: SOFTWARE

The data analyses presented in this paper have been performed using the **mboost** add-on package to the R system of statistical computing. The theoretical ingredients of boosting algorithms, such as loss functions and their negative gradients, base learners and internal stopping criteria, find their computational counterparts in the **mboost** package. Its implementation and user-interface reflect our statistical perspective of boosting as a tool for estimation in structured models. For example, and extending the reference implementation of tree-based gradient boosting from the **gbm** package [74], **mboost** allows to fit potentially high-dimensional linear or smooth additive models, and it has methods to compute degrees of freedom which in turn allow for the use of information criteria such as AIC or BIC or for estimation of variance. Moreover, for high-dimensional (generalized) linear models, our implementation is very fast to fit models even when the dimension of the predictor space is in the ten-thousands.

The Family function in **mboost** can be used to create an object of class *boost_family* implementing the negative gradient for general surrogate loss functions. Such an object can later be fed into the fitting procedure of a linear or additive model which optimizes the corresponding empirical risk (an example is given in Section 5.2). Therefore, we are not limited to already implemented boosting algorithms, but can easily set up our own boosting procedure by implementing the negative gradient of the surrogate loss function of interest.

Both the source version as well as binaries for several operating systems of the **mboost** [43] package are freely available from the Comprehensive R Archive Network (http://CRAN.R-project.org). The reader can install our package directly from the R prompt via

```
R> install.packages("mboost",
                    dependencies =
                    TRUE)
R> library("mboost")
```

All analyses presented in this paper are contained in a package vignette. The rendered output of the analyses is available by the R-command

```
R> vignette("mboost_illustrations",
            package = "mboost")
```

whereas the R code for reproducibility of our analyses can be assessed by

```
R> edit(vignette
          ("mboost_illustrations",
           package = "mboost"))
```

There are several alternative implementations of boosting techniques available as R add-on packages. The reference implementation for tree-based gradient boosting is **gbm** [74]. Boosting for additive models based on penalized B-splines is implemented in **GAMBoost** [9, 84].

## APPENDIX A.2: DERIVATION OF BOOSTING HAT MATRICES

*Derivation of* (7.3). The negative gradient is

$$-\frac{\partial}{\partial f}\rho(y,f) = 2(y-p),$$

$$p = \frac{\exp(f)}{\exp(f) + \exp(-f)}.$$

Next, we linearize $\hat{p}^{[m]}$: we denote $\hat{p}^{[m]} = (\hat{p}^{[m]}(X_1), \ldots, \hat{p}^{[m]}(X_n))^\top$ and analogously for $\hat{f}^{[m]}$. Then,

$$\hat{p}^{[m]} \approx \hat{p}^{[m-1]} + \frac{\partial p}{\partial f}\Big|_{f=\hat{f}^{m-1}}(\hat{f}^{[m]} - \hat{f}^{[m-1]})$$
(A.1)
$$= \hat{p}^{[m-1]} + 2W^{[m-1]}\nu\mathcal{H}^{(\hat{\mathcal{S}}_m)}2(\mathbf{Y} - \hat{p}^{[m-1]}),$$

where $W^{[m]} = \text{diag}(\hat{p}(X_i)(1-\hat{p}(X_i)); 1 \leq i \leq n)$. Since for the hat matrix, $\mathcal{B}_m \mathbf{Y} = \hat{p}^{[m]}$, we obtain from (A.1)

$$\mathcal{B}_1 \approx \nu 4W^{[0]}\mathcal{H}^{\hat{\mathcal{S}}_1},$$
$$\mathcal{B}_m \approx \mathcal{B}_{m-1} + \nu 4W^{[m-1]}\mathcal{H}^{\hat{\mathcal{S}}_m}(I - \mathcal{B}_{m-1}) \quad (m \geq 2),$$

which shows that (7.3) is approximately true.

*Derivation of formula* (7.4). The arguments are analogous to those for the binomial case above. Here, the negative gradient is

$$-\frac{\partial}{\partial f}\rho(y,f) = y - \lambda, \quad \lambda = \exp(f).$$

When linearizing $\hat{\lambda}^{[m]} = (\hat{\lambda}^{[m]}(X_1), \ldots, \hat{\lambda}^{[m]}(X_n))^\top$ we get, analogously to (A.1),

$$\hat{\lambda}^{[m]} \approx \hat{\lambda}^{[m-1]} + \frac{\partial \lambda}{\partial f}\Big|_{f=\hat{f}^{m-1}}(\hat{f}^{[m]} - \hat{f}^{[m-1]})$$
$$= \hat{\lambda}^{[m-1]} + W^{[m-1]}\nu\mathcal{H}^{(\hat{\mathcal{S}}_m)}(\mathbf{Y} - \hat{\lambda}^{[m-1]}),$$

where $W^{[m]} = \text{diag}(\hat{\lambda}(X_i)); 1 \leq i \leq n$. We then complete the derivation of (7.4) as in the binomial case above.

## ACKNOWLEDGMENTS

We would like to thank Axel Benner, Florian Leitenstorfer, Roman Lutz and Lukas Meier for discussions and detailed remarks. Moreover, we thank four referees, the editor and the executive editor Ed George for constructive comments. The work of T. Hothorn was supported by Deutsche Forschungsgemeinschaft (DFG) under grant HO 3242/1-3.